\documentclass{article}
\usepackage{graphics}
\usepackage{rotating}

\begin{document}

\title{The Network Structure of Mathematical Knowledge According to the
Wikipedia, MathWorld, and DLMF Online Libraries}

\author{Flavio~B.~Gonzaga$^{1,2}$\\
Valmir~C.~Barbosa$^{1,}$\thanks{Corresponding author
(valmir@cos.ufrj.br).}$\mbox{\ }$\\
Geraldo~B.~Xex\'eo$^1$\\
\\
$^1$Programa de Engenharia de Sistemas e Computa\c c\~ao, COPPE\\
Universidade Federal do Rio de Janeiro\\
Caixa Postal 68511\\
21941-972 Rio de Janeiro - RJ, Brazil\\
\\
$^2$N\'ucleo de Ci\^encia da Computa\c c\~ao\\
Universidade Federal de Alfenas\\
37130-000 Alfenas - MG, Brazil}

\date{}

\maketitle

\begin{abstract}
We study the network structure of Wikipedia (restricted to its mathematical
portion), MathWorld, and DLMF. We approach these three online mathematical
libraries from the perspective of several global and local network-theoretic
features, providing for each one the appropriate value or distribution, along
with comparisons that, if possible, also include the whole of the Wikipedia or
the Web. We identify some distinguishing characteristics of all three libraries,
most of them supposedly traceable to the libraries' shared nature of relating to
a very specialized domain. Among these characteristics are the presence of a
very large strongly connected component in each of the corresponding directed
graphs, the complete absence of any clear power laws describing the distribution
of local features, and the rise to prominence of some local features (e.g.,
stress centrality) that can be used to effectively search for keywords in the
libraries.

\bigskip
\noindent
\textbf{Keywords:} Online mathematical libraries, Wikipedia, MathWorld, DLMF,
complex networks, text search.
\end{abstract}

\newpage
\section{Introduction}\label{intr}

Until a few decades ago, before it became commonplace to search the Web for
information and knowledge, people desiring quick access to some mathematical
concept or formula used to resort to printed encyclopedias or handbooks, such as
the compilation by Abramowitz and Stegun \cite{as65}, the more specialized
tables put together by Gradshteyn and Ryzhik \cite{gr00}, or still others
\cite{i93,bsmm04}. Of such volumes, the undisputed citations champion seems to
be Abramowitz and Stegun's \cite{bclo11}, whose work has since been methodically
expanded \cite{l03} into a NIST-sponsored publication \cite{olbc10}.

Lately, though, the situation has become, if anything, more complex. For, while
those printed works continue to be used and cited widely and their ranks
continue to be enlarged by the addition of new works of a similar genre
\cite{gbl08}, the premier source, at least for a first approach, has undoubtedly
become the Web. In fact, it seems safe to state that most mathematics-related
queries on Google return
Wikipedia\footnote{\texttt{http://en.wikipedia.org/wiki/Portal:Mathematics}.}
or Wolfram MathWorld\footnote{\texttt{http://mathworld.wolfram.com}.} pages as
prominently ranked. As mentioned, however, printed and online material still
coexist and, curiously, movement has taken place in both directions: while in
one direction MathWorld material has found its way into Weisstein's
encyclopedia \cite{w02}, in the other the NIST volume has been turned into the
Digital Library of Mathematical Functions,
DLMF.\footnote{\texttt{http://dlmf.nist.gov}.}

Here we aim to explore the structure of mathematical knowledge as reflected in
these three online libraries. By ``structure'' we do not mean the organization
of material into the many mathematical areas and subareas. Nor do we mean the
coalescence of all deduction chains that is behind all of mathematics and
inherently amounts to an acyclic directed graph \cite{c02}, i.e., one with no
directed cycles. We mean, rather, the no longer acyclic directed graphs that
reflect all the cross-referencing that took place as those libraries were
created by several collaborators (and still takes place as the libraries
evolve). Exploring their graph structures from the perspective of such
hypertextual interconnections amounts to applying some of the complex-network
notions and metrics developed during the past fifteen years or so, much as has
been done so successfully to various other fields \cite{bs03,nbw06,bkm09}.

It also amounts to a chance to globally view all the material compiled into each
library and inquire, from a network-theoretic perspective, what traces remain,
if any, as telltale signs of the essentially very distinct methods of
construction employed to build them, all of a collaborative nature but
supposedly more and more controlled as we move from Wikipedia to MathWorld and
then to DLMF. In our analyses we use several frequency data, of both a
network-wide nature as well as node-related, aiming not only to describe the
libraries' properties as such data reveal them, but also to discover how these
properties relate to the libraries' robustness in the face of accidental or
intentional loss of material and to their ease of search in response to text
queries.

What has turned up is a collection of results that both sets the three
mathematical libraries apart from the wider English-language Wikipedia and from
the much wider Web, and at the same time groups the three libraries together
insofar as they share important properties. Some of the most significant
results include the characteristic that a very large fraction of each library's
pages are packed together in the sense of mutual reachability; the presence of
clear signs that all three libraries result from decisions regarding the
deployment of links that are leveraged by technical knowledge (rather than, say,
some nontechnical measure of a page's relevance, such as popularity); and the
discovery of successful criteria for guiding text search within the libraries'
pages that differ significantly from those most commonly used (e.g., by Google).

We proceed in the following manner. First, in Section~\ref{graphs}, we introduce
the five directed graphs that we use in all analyses (two for Wikipedia, two for
MathWorld, one for DLMF) and also some basic notation. We then move,
respectively in Sections~\ref{global} and~\ref{local}, to a study of these
graphs' global and local network-theoretic features. Section~\ref{local-scc} is
dedicated to an analysis of the five graphs' robustness when nodes are lost
either as a result of some random process or as a deterministic function of the
graphs' local features. We continue with Section~\ref{local-search}, where we
investigate the effect of such features in the ranking of nodes when responding
to text queries. We conclude in Section~\ref{concl}.

\section{Five directed graphs}\label{graphs}

In all three libraries it is possible to reach the technical-content pages by
navigating through a hierarchy of specialized subdivisions from the main portal
(the so-called category pages). Once the content pages are reached, further
navigation is possible through the links that lead from one such page to
another. Each of the directed graphs with which we work has a node for each
content page and directed edges that reflect inter-page links. In all cases,
links leading from a page to itself are ignored when building the graph, so no
self-loops exist. Similarly, should multiple links exist from a page to another,
only one edge is created in the graph between the corresponding nodes.

In the case of Wikipedia and MathWorld, links can be categorized into those
appearing in a page's main text and those that are given in the page's ``See
also'' section when it exists. We perceive these two link types as playing
entirely different roles. While in-text links are generally meant to clarify
some of the terms used in the page, being therefore meant for quick side lookups
before continuing on the main text, See-also links are used to point to pages
where related material is to be found. For this reason, we use two different
graphs for each of Wikipedia and MathWorld. They both have the same node set,
but their edge sets differ, one reflecting in-text as well as See-also links,
the other reflecting See-also links only.

The case of DLMF requires no such special treatment. Although its pages, too,
contain special, ``Referenced by'' links, such links are simply antiparallel
versions of the library's non-Referenced-by links. That is, page $a$ contains a
non-Referenced-by link to page $b$ if page $b$ contains a Referenced-by link to
page $a$. Referenced-by links in DLMF are therefore redundant as far as building
its directed graph is concerned. They are for this reason ignored.

These observations amount to five different graphs with which to work, as
summarized in Table~\ref{table-graphs}. In the table, for each of the libraries
and, when applicable, taking See-also links into account, we give the time frame
within which the content pages were downloaded and the notation we use to refer
to the corresponding graph.

\begin{table}
\centering
\caption{Online libraries and corresponding directed graphs.}
\label{table-graphs}
\begin{tabular}{lll}
\hline
Library & Download period & Directed graph \\
\hline
Wikipedia & September 2010 & $W$ \\
Wikipedia, See-also links & September 2010 & $W'$ \\
MathWorld & August 2009 & $M$ \\
MathWorld, See-also links & August 2009 & $M'$ \\
DLMF & September 2010 & $D$ \\
\hline
\end{tabular}

\end{table}

Some additional basic notation to be used throughout is the following. Given the
graph under consideration, we let $n$ stand for its number of nodes and $m$ for
its number of edges. For node $i$, $I_i$ is its set of in-neighbors (nodes from
which edges are directed toward $i$) and $O_i$ its set of out-neighbors (nodes
toward which edges are directed from $i$). Its in-degree is
$\delta_i^+=\vert I_i\vert$, its out-degree is $\delta_i^-=\vert O_i\vert$, and
its number of neighbors when edge directions are disregarded (henceforth
referred to simply as its degree) is
$\delta_i=\vert I_i\cup O_i\vert\le\delta_i^++\delta_i^-$. Clearly, it holds
that $\max\{\delta_i^+,\delta_i^-\}\le\delta_i$. For any two nodes $i$ and $j$,
$d_{ij}$ is the distance from $i$ to $j$, that is, the number of edges on a
shortest directed path leading from $i$ to $j$. If none exists, then
$d_{ij}=\infty$. We let $R_i$ be the set of nodes $j$ such that
$0<d_{ij}<\infty$. Note that $R_i=\emptyset$ if and only if node $i$ is a sink,
i.e., $O_i=\emptyset$.

\section{Global features}\label{global}

We give six global features for each graph. The first two are straightforward
and provide simple relationships between the graph's number of nodes, $n$, and
its number of edges, $m$. The first one is simply the graph's mean in-degree,
denoted by $\delta^+$ and given by
\begin{equation}
\delta^+
=\frac{1}{n}\sum_i\delta_i^+
=\frac{m}{n}
\end{equation}
(necessarily equal to the graph's mean out-degree). The second feature is the
graph's mean degree. Denoting it by $\delta$, we have
\begin{equation}
\delta^+
\le\delta
=\frac{1}{n}\sum_i\delta_i
\le\frac{1}{n}\sum_i(\delta_i^++\delta_i^-)
=2\delta^+.
\end{equation}
Both $\delta^+$ and $\delta$ work as indicators of the graph's edge density
relative to its number of nodes. The value of $\delta$, in particular, may swing
toward either of its bounds, $\delta^+$ and $2\delta^+$, indicating in the
former case that every edge's antiparallel counterpart is also present in the
graph and in the latter case that none is. On average, then, the fraction of
$\delta$ corresponding to antiparallel edge pairs is given by
$(2\delta^+-\delta)/\delta=2\delta^+/\delta-1$.

Our next global feature is the fraction $S$ of $n$ that corresponds to the nodes
inside the graph's largest strongly connected component (GSCC henceforth, where
$G$ is for ``giant''). A strongly connected component is either a singleton
whose only member, say node $i$, is such that $i\notin R_j$ for every node
$j\in R_i$ (no directed path exists back from any node that can be reached from
$i$ through a directed path), or a larger set that is maximal with respect to
the property that $j\in R_i$ for any two of its members $i$ and $j$ such that
$j\neq i$. In the latter case, then, a directed path exists between any two
distinct nodes inside the strongly connected component. Informally, the value of
$S$ can be regarded as an indication of the network's ``degree of acyclicity.''
If the graph is acyclic, then all its strongly connected components are
singletons and $S=1/n$. The other extreme corresponds to the case in which all
nodes are in the GSCC, so $S=1$.

The fourth and fifth global features are both related to classifying a graph
vis-\`{a}-vis the so-called small-world criteria \cite{ws98,ajb99}, namely small
distances and large transitivity. We address the first criterion by computing
the average distance between any two distinct nodes, so long as only finite
distances are considered. We denote this average by $\ell$, which is then such
that
\begin{equation}
\ell=\frac{1}{N}\sum_i\sum_{j\in R_i}d_{ij},
\end{equation}
where $N$ is the number of $i,j$ pairs contributing to the double summation. As
for the second criterion, that of transitivity, we follow the usual trend of
disregarding edge directions and computing the resulting graph's clustering
coefficient in its most common formulation \cite{n10}. If $C$ is the clustering
coefficient, then this formulation lets $C=3t/T$, where both $t$ and $T$ refer
to node triples in the graph, e.g., $i,j,k$. The value of $t$ is meant to
reflect the number of triangles in the graph, that is, those triples in which an
edge connects $i$ and $j$, another connects $j$ and $k$, and yet another
connects $i$ and $k$. The value of $T$, on the other hand, counts the triples
that are arranged as three-node (two-edge) paths. The factor $3$ in the
numerator of the ratio defining $C$ reflects the fact that there are three
triples of the latter type for each triangle in the graph. It follows that
$0\le C\le 1$ (no transitivity through full transitivity). In our analysis of
each graph's clustering coefficient $C$, we present it side-by-side with the
value it would have if every node $i$ continued to have the same degree
$\delta_i$ but the connections were made at random \cite{n10}. This value,
denoted by $C'$, is given by
\begin{equation}
C'=\frac{(\delta^{(2)}-\delta)^2}{n\delta^3},
\end{equation}
where $\delta^{(2)}=(1/n)\sum_i\delta_i^2$.

Our last global feature is in fact a series of four assortativity coefficients.
Each one is the Pearson correlation coefficient of two length-$m$ sequences of
numbers. If $\alpha_1,\alpha_2,\ldots,\alpha_m$ and
$\beta_1,\beta_2,\ldots,\beta_m$ are the sequences, $\mu_\alpha$ and $\mu_\beta$
are the corresponding means, and $\sigma_\alpha$ and $\sigma_\beta$ are the
corresponding standard deviations, this coefficient is
\begin{equation}
r_{\alpha,\beta}=\frac
{(1/m)\sum_e\alpha_e\beta_e-\mu_\alpha\mu_\beta}
{\sigma_\alpha\sigma_\beta}.
\end{equation}
The original assortativity coefficient is obtained by letting
$\alpha_e=\delta_i^-$ and $\beta_e=\delta_j^+$ for $e$ the edge directed from
$i$ to $j$ \cite{n02,n03}. That is, it measures how correlated the out-degrees
of the edges' tail nodes are with the in-degrees of the edges' head nodes. A
shorthand for this formulation is to use out,in in place of $\alpha,\beta$. We
get the other three variations by selecting the other possible combinations
(in,out; out,out; in,in) \cite{ffgp10,ppz11}.

The global features of the graphs in Table~\ref{table-graphs} are shown in
Tables~\ref{table-global1} and~\ref{table-global2}, which include an additional
row for the directed graph, denoted by $W^+$, that corresponds to the entire
English-language Wikipedia of a relatively recent past \cite{cscbdlc06,zbsd06}.
Table~\ref{table-global1}, moreover, contains one further row for the whole
Web, now based on data from an older past \cite{bkmrrstw00}.\footnote{Slightly
more recent data seem to indicate an $S$ value of roughly $0.33$ for a similarly
sized Web \cite{dllm04}, but no estimate is given for $\ell$.} The corresponding
directed graph is denoted by $W^*$. Not all global features are available for
$W^+$ or $W^*$, as indicated by blank entries in the tables. Graphs are arranged
in Tables~\ref{table-global1} and~\ref{table-global2} in nonincreasing order of
$n$, then in decreasing order of $m$.

\begin{sidewaystable}
\centering
\caption{Global features: mean in- or out-degree ($\delta^+$), mean degree
($\delta$) and the resulting value of $2\delta^+/\delta-1$, fraction of $n$
within the GSCC ($S$), average distance between distinct nodes ($\ell$), and
clustering coefficient ($C$, along with the value, $C'$, it would have if
connections were random).}
\label{table-global1}
\begin{tabular}{lrrrrccrcc}
\hline
Graph & \multicolumn{1}{c}{$n$} & \multicolumn{1}{c}{$m$} & \multicolumn{1}{c}{$\delta^+$} & \multicolumn{1}{c}{$\delta$}  &$2\delta^+/\delta-1$ & $S$ & \multicolumn{1}{c}{$\ell$} & $C$ & $C'$ \\
\hline
$W^*$ & $203\,549\,046$ & $1\,466\,000\,000$ & $7.20$ & & & $0.28$ & $16.18$ & & \\
$W^+$ & $339\,834$ & $5\,278\,037$ & $15.53$ & & & $0.82$ & $4.90$ & & \\
\hline
$W$ & $37\,723$ & $688\,589$ & $18.25$ & $30.62$  & $0.19$ & $0.80$ & $4.11$ & $0.055$ & $7.59\times 10^{-4}$ \\
$W'$ & $37\,723$ & $21\,503$ & $0.57$ & $1.04$  & $0.09$ & $0.02$ & $15.27$ & $0.061$ & $4.96\times10^{-8}$ \\
$M$ & $15\,095$ & $92\,648$ & $6.14$ & $9.72$  & $0.26$ & $0.78$ & $5.32$ & $0.048$ & $5.18\times10^{-4}$ \\
$M'$ & $15\,095$ & $46\,965$ & $3.11$ & $4.45$  & $0.40$ & $0.62$ & $7.45$ & $0.093$ & $1.77\times10^{-4}$ \\
$D$ & $908$ & $7\,527$ & $8.29$ & $12.81$  & $0.29$ & $0.81$ & $3.79$ & $0.062$ & $0.011$ \\
\hline
\end{tabular}

\end{sidewaystable}

\begin{table}
\centering
\caption{Global features: assortativity coefficients.}
\label{table-global2}
\begin{tabular}{lrrrr}
\hline
Graph & \multicolumn{1}{c}{$r_\mathrm{out,in}$} & \multicolumn{1}{c}{$r_\mathrm{in,out}$} & \multicolumn{1}{c}{$r_\mathrm{out,out}$} & \multicolumn{1}{c}{$r_\mathrm{in,in}$} \\
\hline
$W^+$ & $-0.150$ & & & \\
\hline
$W$ & $-0.071$ & $0.075$ & $-0.074$ & $-0.022$ \\
$W'$ & $0.041$ & $0.094$ & $0.070$ & $0.028$ \\
$M$ & $-0.037$ & $-0.018$ & $-0.015$ & $-0.019$ \\
$M'$ & $-0.054$ & $-0.031$ & $-0.058$ & $-0.036$ \\
$D$ & $-0.169$ & $0.006$ & $-0.053$ & $-0.043$ \\
\hline
\end{tabular}

\end{table}

The data shown in Table~\ref{table-global1} indicate that edge density relative
to the number of nodes, as given by $\delta^+$, has the same order of magnitude
for most graphs, the exception being $W'$, the Wikipedia graph based exclusively
on See-also links, whose $\delta^+$ value is one order of magnitude lower.
Wikipedia contributors to the mathematical pages, therefore, seem to deploy
See-also links considerably less methodically than those who contribute to
MathWorld. It is also worth noting that the five mathematics-related graphs have
fairly different values for the ratio $2\delta^+/\delta-1$, pointing at $W'$ as
the graph with the fewest antiparallel edge pairs contributing to degrees on
average, and to $M'$, the MathWorld graph based on See-also links, as having the
most. Once again, then, MathWorld contributors appear more meticulous at
providing cross-referencing information of the See-also form.

One of the most striking contrasts in Table~\ref{table-global1} concerns the
value of $S$, the size of the graph's GSCC relative to $n$. While for the Web
graph $W^*$ our current best estimate places about $28\%$ of the nodes inside
the GSCC, for the Wikipedia graph $W^+$ and most of the mathematical-library
graphs we have been considering the GSCC encompasses substantially more nodes
(between $62$ and $82\%$). The exception, once again, occurs on account of graph
$W'$, whose GSCC is sized at a mere $2\%$ of the nodes, and which as we have
noted is only very sparsely interconnected by the See-also links.

The remaining data in Table~\ref{table-global1} refer to $\ell$ and to $C$, a
graph's average path length (in the directed sense) and clustering coefficient
(in the undirected sense), respectively. We first note that, for six of the
seven graphs, $\ell$ is proportional to $\ln n$ by a constant of the order of
$10^{-1}$, the exception being $W'$, for which the proportionality constant is
roughly $1.45$ (this comes from the substantially larger distances in comparison
to $W$, as expected from the substantially lower $\delta^+$ value). In all
cases, however, distances are on average very small given the value of $n$, so
all seven graphs qualify as small-world structures. Moreover, as is usually but
not always the case \cite{n10}, in all five mathematics-related graphs the value
of $C$ is noticeably larger than that of $C'$. In fact, except for the DLMF
graph $D$, $C$ surpasses $C'$ by a factor of at least two orders of magnitude.
The construction of $D$, which has $C\approx 5.64C'$, seems to have been guided
by forces that prevent the formation of triangles more than they do in the other
four cases. One possible explanation is that, in comparison to Wikipedia or
MathWorld, each DLMF page contains substantially more material, which in fact is
reflected in the low number of nodes of graph $D$.

Table~\ref{table-global2} contains all four assortativity coefficients for all
five mathematics-related graphs and $W^+$, the unrestricted Wikipedia graph. The
vast majority of all values is of the order of $10^{-2}$ at most, being
therefore sufficiently near zero for the sequences involved to be taken as
uncorrelated. In general this is indicative either of a random pattern of
connections (which is not the case) or that criteria for edge deployment are at
work that make no reference whatsoever to in- or out-degrees (which is more
plausibly the case). Curiously, though, the same holds also for the only two
exceptions, $W^+$ and $D$, for which the moderately negative but nonnegligible
value of $r_\mathrm{out,in}$ is suggestive that in these two graphs connections
are effected in such a way that promotes a small but noticeable degree of
disassortative mixing of tail nodes' out-degrees with head nodes' in-degrees.
That is, there is a slight tendency of nodes with larger (smaller) out-degrees
to connect out to nodes with smaller (larger) in-degrees. This tendency is
quantified very similarly by $r_\mathrm{out,in}$ for both $W^+$ and $D$
($-0.150$ in the former case, $-0.169$ in the latter). Perhaps the
aforementioned fact that the typical DLMF page contains more material than the
other libraries' pages somehow makes $D$ resemble $W^+$ in this one aspect.

\section{Local features}\label{local}

Presenting a graph's local features requires that we value each feature of
interest for each node and then provide some probability distribution of that
feature over the entire graph. In this section we work with the feature's
complementary cumulative distribution (CCD henceforth), denoted by $F(z)$ for
an admissible feature value $z$, which is the probability that a randomly chosen
node has a feature value that surpasses $z$. We compute $F(z)$ as the fraction
of $n$ representing the nodes for which the feature is valued beyond $z$.
Clearly, if for a graph the feature in question is never valued beyond $Z$, then
$F(z)=0$ for all $z\ge Z$.

The most widely studied local features are a node's in-degree, out-degree, and
degree. Not only have they been measured in a variety of domains, but knowledge
of how they are distributed can be used in the study of many other network
properties \cite{nsw01}. These features are the first three we study, as
characterizations of in- and out-degrees have over the years led to important
discoveries regarding the Web and the unrestricted Wikipedia. Specifically, we
know from at least two independent sources operating on different data that
in-degrees in the Web graph (our $W^*$ graph in one case, a different version in
the other) are distributed according to a power law
\cite{ajb99,bkmrrstw00,dllm04}. That is, the probability that a randomly chosen
node has in-degree $k>0$ is proportional to $k^{-\alpha}$ (so the corresponding
CCD is approximately proportional to $k^{1-\alpha}$) for $\alpha\approx 2.1$.
Similar power laws have also been reported for the graph's out-degrees, but in
this case there seems to be some disagreement \cite{dllm04}. As for the
Wikipedia graph, $W^+$, its in-degree, out-degree, and degree distributions have
all been found to follow power laws, of exponents $-2.21$, between $-2.65$ and
$-2$, and $-2.37$, respectively \cite{cscbdlc06,zbsd06}. Power laws are
inherently scale-free \cite{n05}, and their appearance in graphs such as $W^*$
and $W^+$ has been explained particularly well by the mechanism of edge
deployment known as preferential attachment \cite{p76,ba99,baj00,cscbdlc06}.

The additional local features that we consider are the ones given in
Table~\ref{table-local}. Four of them ($B_i$, $S_i$, $C_i$, and $G_i$) are
measures of node $i$'s centrality in the graph, being therefore related to
shortest directed paths in which $i$ participates in some way. The remaining
three are related to search mechanisms on the Web. They are measures of how node
$i$ qualifies as a hub ($y_i$) or an authority ($x_i$) in the HITS
(Hyperlink-Induced Topic Search) mechanism, and the node's page rank ($\rho_i$),
which underlies Google searches.

\begin{sidewaystable}
\centering
\caption{Additional local features for node $i$. $\sigma_{jk}$ is the number of
shortest directed paths that lead from $j$ to $k$, while $\sigma_{jk}(i)$ counts
only the ones that go through $i$.}
\label{table-local}
\begin{tabular}{lll}
\hline
Designation & Formula & Reference(s) \\
\hline
&&\\[-7pt]
Betweenness centrality &
$\displaystyle{B_i=\sum_{j\neq i}\sum_{{k\neq i}\atop{k\in R_j}}
\frac{\sigma_{jk}(i)}{\sigma_{jk}}}$ &
\cite{a71,f77}\\[22pt]
Stress centrality &
$\displaystyle{S_i=\sum_{j\neq i}\sum_{{k\neq i}\atop{k\in R_j}}
\sigma_{jk}(i)}$ &
\cite{s53}\\[22pt]
Closeness centrality &
$\displaystyle{C_i=\left\{\begin{array}{ll}
\displaystyle
\frac{1}{\sum_{j\in R_i}d_{ij}},&\mbox{if $R_i\neq\emptyset$}\\\hfill 0,&\mbox{otherwise}
\end{array}\right.}$ &
\cite{s66}\\[22pt]
Graph centrality &
$\displaystyle{G_i=\left\{\begin{array}{ll}
\displaystyle
\frac{1}{\max_{j\in R_i}d_{ij}},&\mbox{if $R_i\neq\emptyset$}\\\hfill 0,&\mbox{otherwise}
\end{array}\right.}$ &
\cite{hh95}\\[22pt]
HITS update rule for hubs &
$\displaystyle{y_i:=\sum_{j\in O_i}x_j}$ &
\cite{k99}\\[22pt]
HITS update rule for authorities &
$\displaystyle{x_i:=\sum_{j\in I_i}y_j}$ &
\cite{k99}\\[22pt]
Page-rank update rule with damping factor $0.85$ &
$\displaystyle{\rho_i:=0.15+0.85\sum_{j\in I_i}\frac{\rho_j}{\delta_j^-}}$ &
\cite{bp98}\\[15pt]
\hline
\end{tabular}

\end{sidewaystable}

The centrality features can be computed through variations of a well-known
algorithm \cite{ub01}, and similarly the other three, though requiring iterative
updates for convergence. In the case of the HITS-related features, first every
$x_i$ and $y_i$ is initialized to $1$. Then the $x_i$'s and $y_i$'s are
alternately updated via the rules given in Table~\ref{table-local}. The updating
of the $x_i$'s is followed by a normalization of the resulting values so that
$\sum_ix_i^2=1$, which is achieved by dividing each $x_i$ by the Euclidean norm
of the vector of components $x_1,x_2,\ldots,x_n$. The updating of the $y_i$'s
proceeds similarly. After convergence, all features are normalized so that
$\sum_ix_i=\sum_iy_i=1$. As for the page-rank feature, once again every $\rho_i$
is initialized to $1$ and the update rule given in Table~\ref{table-local} is
iteratively applied until convergence, at which time all $\rho_i$'s are
normalized so that $\sum_i\rho_i=1$. For the two HITS-related features and page
rank, our criterion for convergence has been that, for all nodes, the two latest
feature values differ from each other by some quantity in the interval
$[-10^{-16},10^{-16}]$.

\begin{figure}[p]
\centering
\scalebox{0.65}{\includegraphics{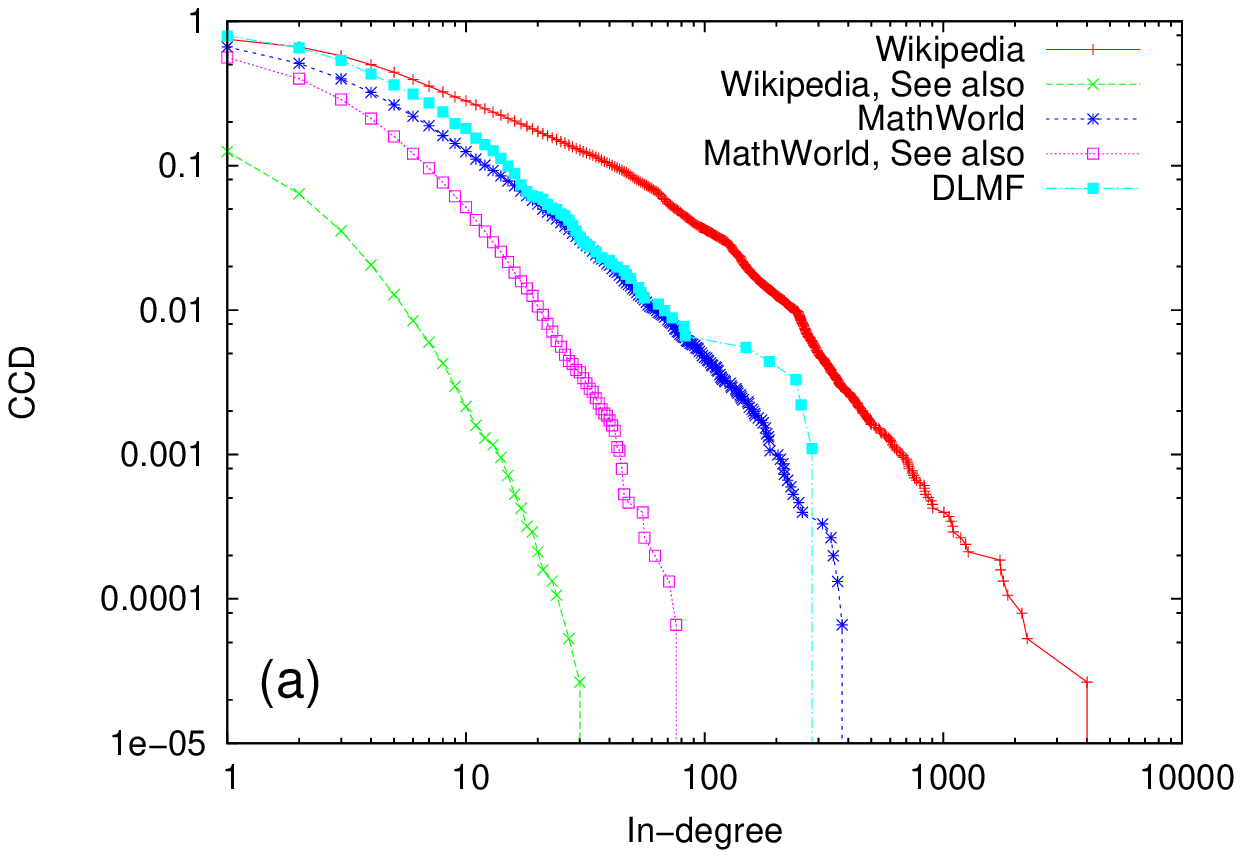}}
\scalebox{0.65}{\includegraphics{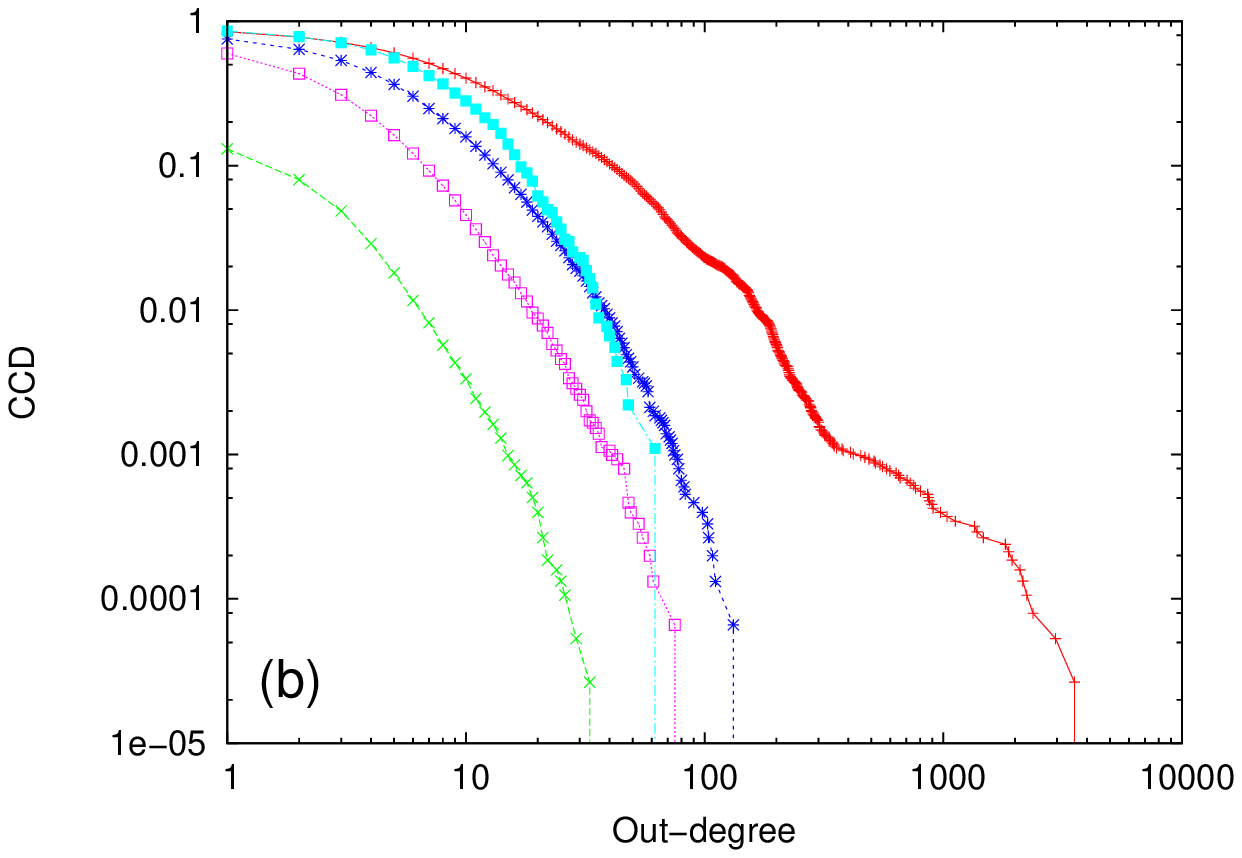}}
\scalebox{0.65}{\includegraphics{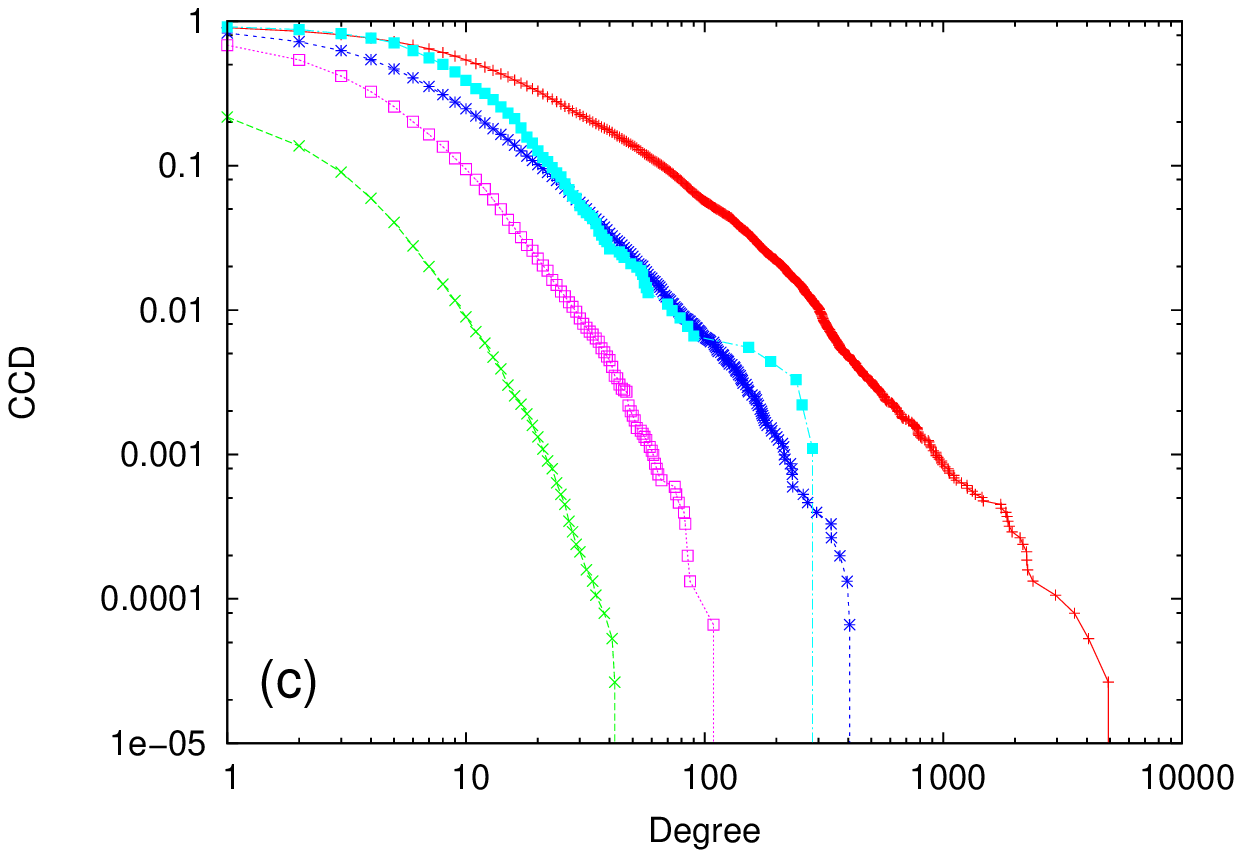}}
\caption{CCD plots for the $\delta_i^+$ (a), $\delta_i^-$ (b), and $\delta_i$
(c) values of $W$ (Wikipedia), $W'$ (Wikipedia, See also), $M$ (MathWorld), $M'$
(MathWorld, See also), and $D$ (DLMF).}
\label{fig-degrees}
\end{figure}

CCD plots for the local features are given in Figure~\ref{fig-degrees}
(in-degree, out-degree, and degree), Figure~\ref{fig-centralities}
(centralities), and Figure~\ref{fig-searches} (hub, authority, and page rank).
One striking characteristic they all share is that no feature of any library
seems to be expressible as a clear power law for any significant number of
orders of magnitude. For example, although we have found the in-degree CCD for
DLMF to be given approximately by a power-law of $\alpha=2.47$, this seems
reasonable only for one order of magnitude (roughly between $10$ and $100$).
In the case of Figure~\ref{fig-degrees}, in particular, this widespread absence
of a power law works to confirm the expectation that, in such a specialized
domain as the five libraries', it is expertise, rather than some
popularity-based criterion such as preferential attachment, that guides the
establishment of connections.

\begin{figure}[p]
\centering
\scalebox{0.65}{\includegraphics{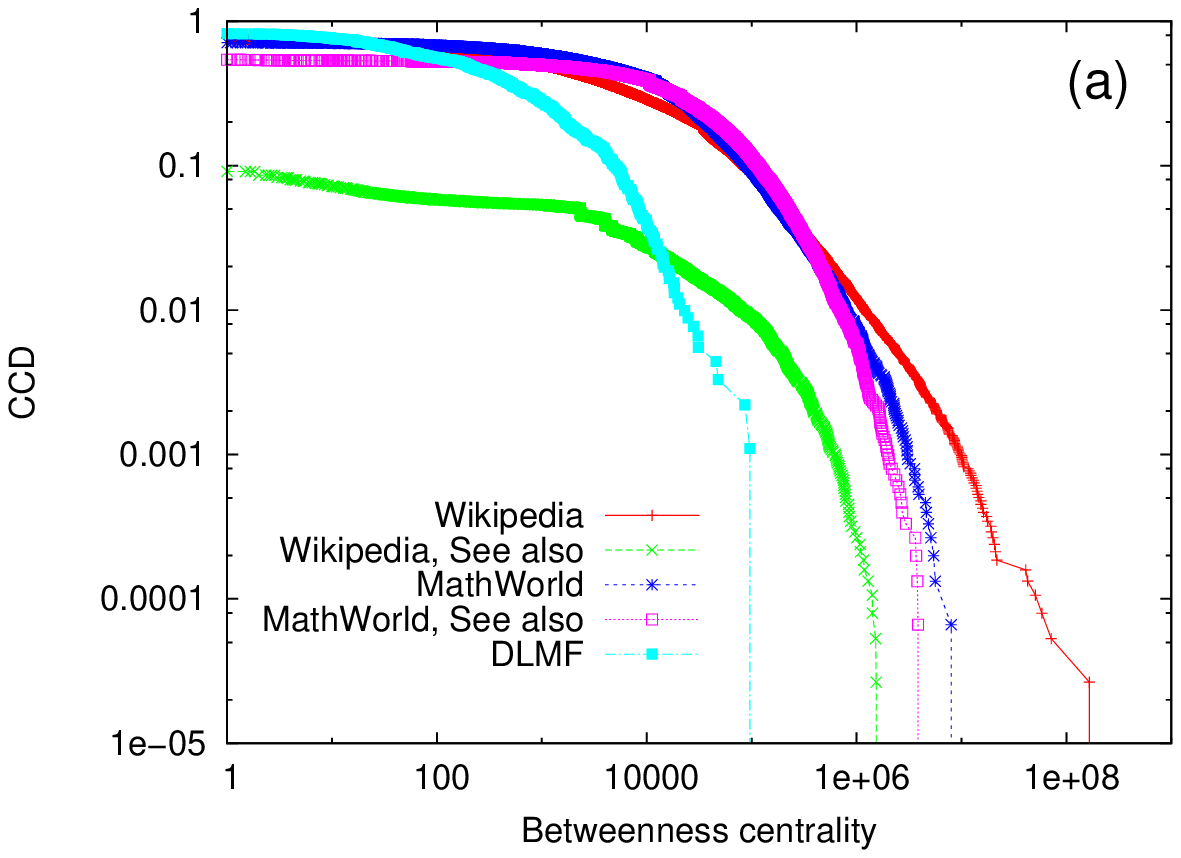}}
\scalebox{0.65}{\includegraphics{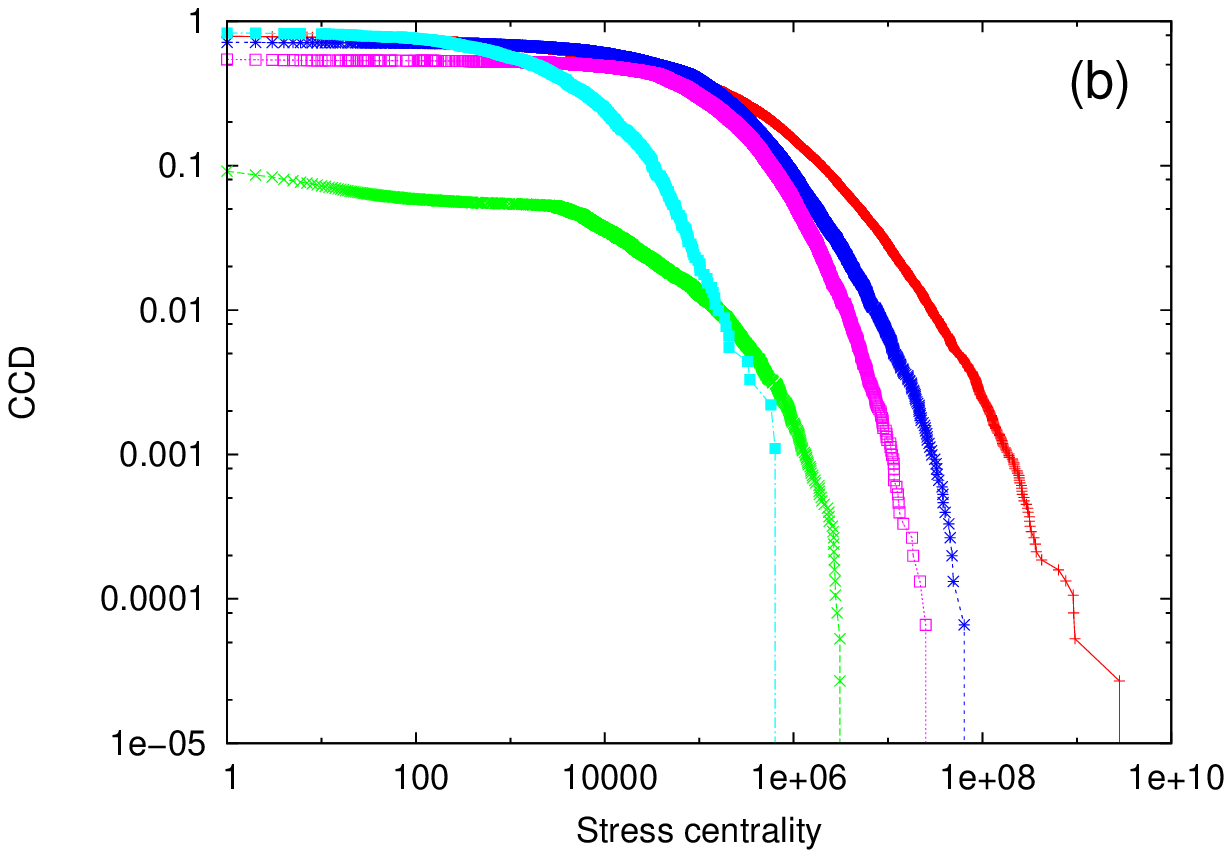}}
\scalebox{0.65}{\includegraphics{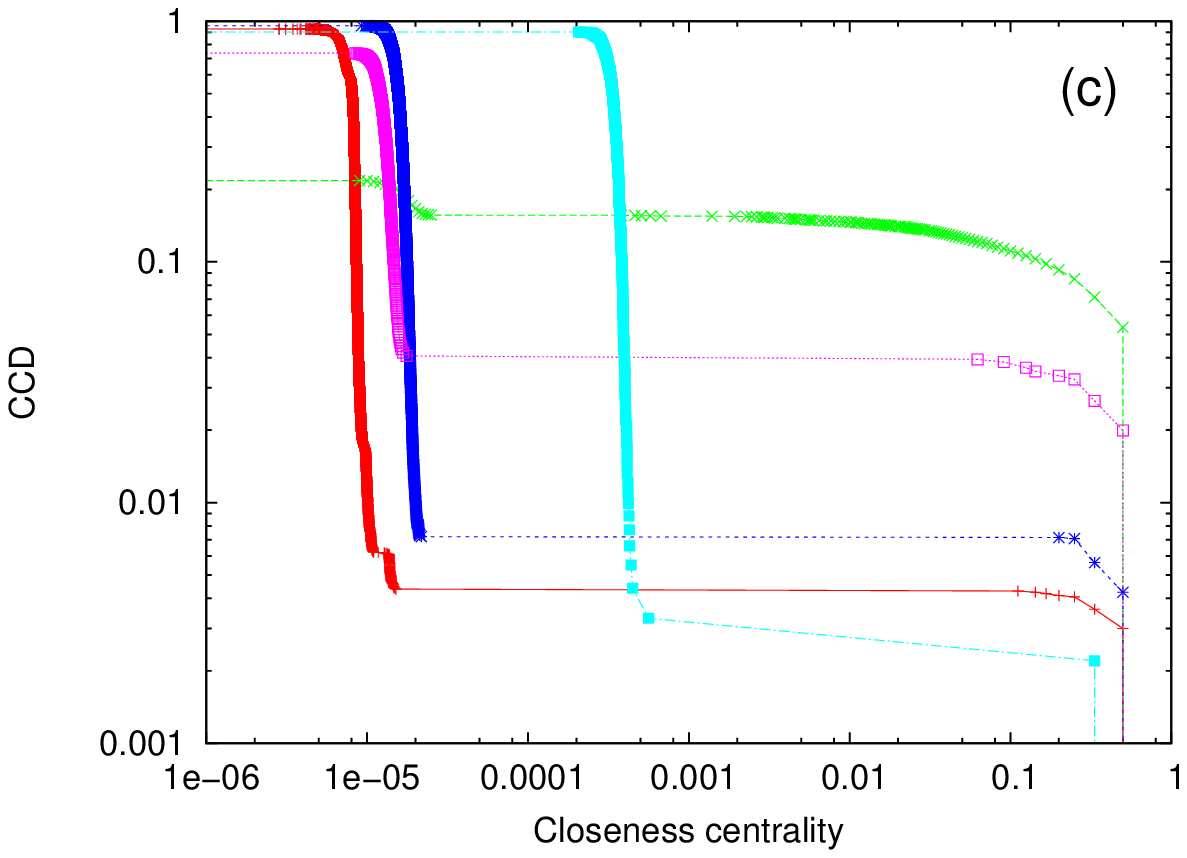}}
\caption{CCD plots for the $B_i$ (a), $S_i$ (b), $C_i$ (c), and $G_i$ (d) values
of $W$ (Wikipedia), $W'$ (Wikipedia, See also), $M$ (MathWorld), $M'$
(MathWorld, See also), and $D$ (DLMF).}
\label{fig-centralities}
\end{figure}

\addtocounter{figure}{-1}
\begin{figure}
\centering
\scalebox{0.65}{\includegraphics{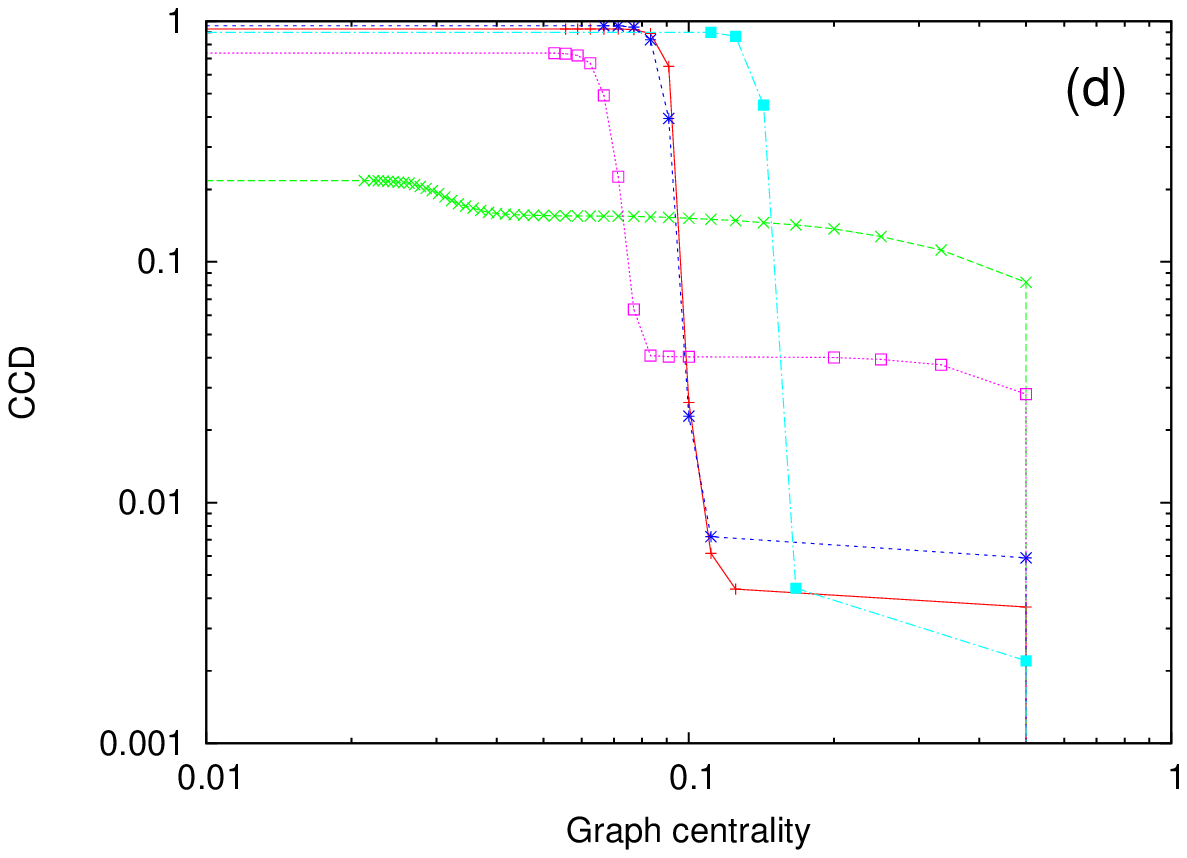}}
\caption{Continued.}
\end{figure}

\begin{figure}[p]
\centering
\scalebox{0.65}{\includegraphics{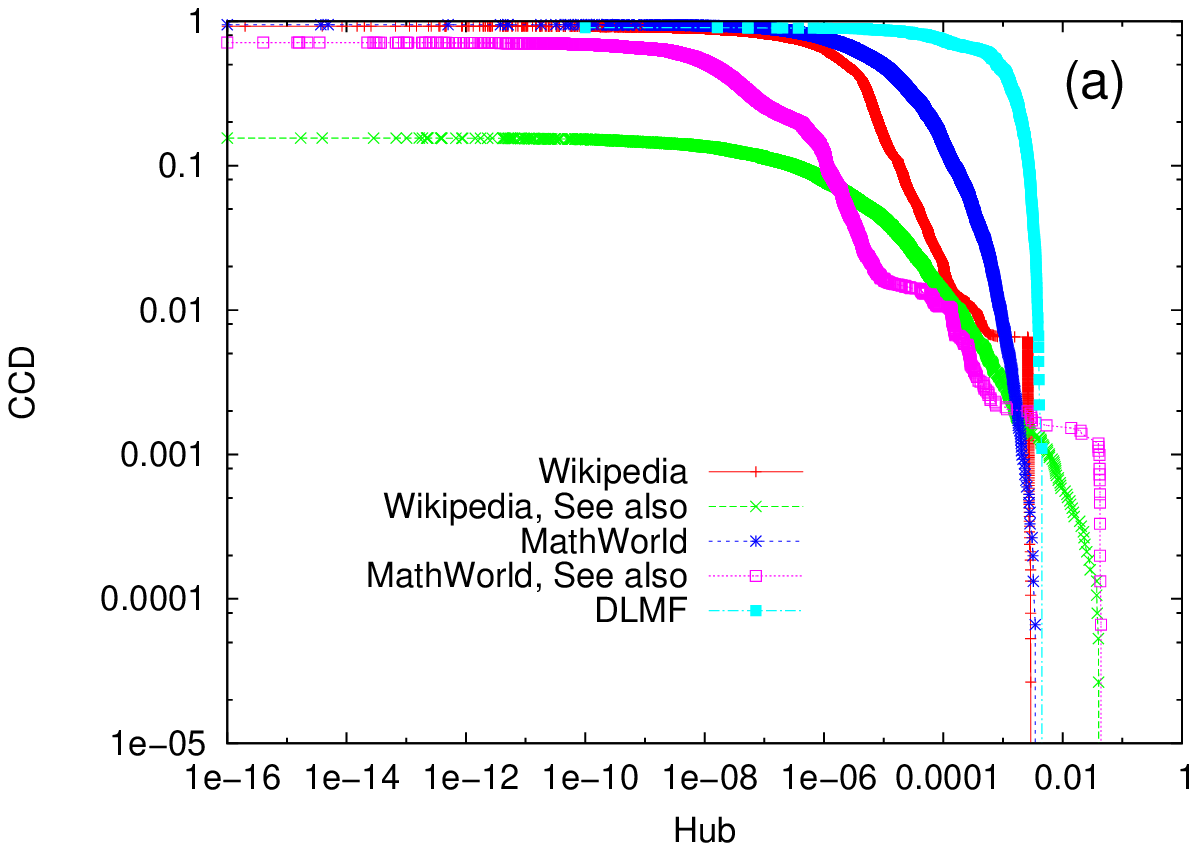}}
\scalebox{0.65}{\includegraphics{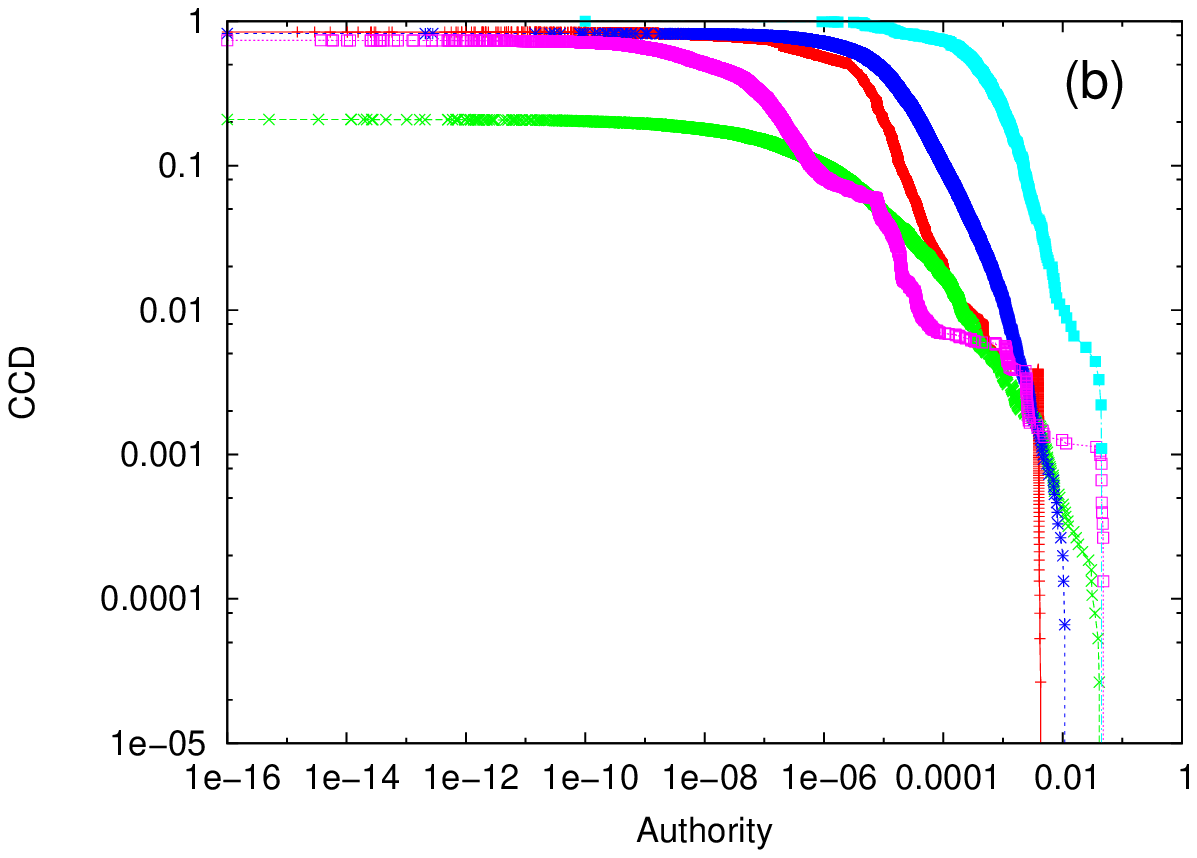}}
\scalebox{0.65}{\includegraphics{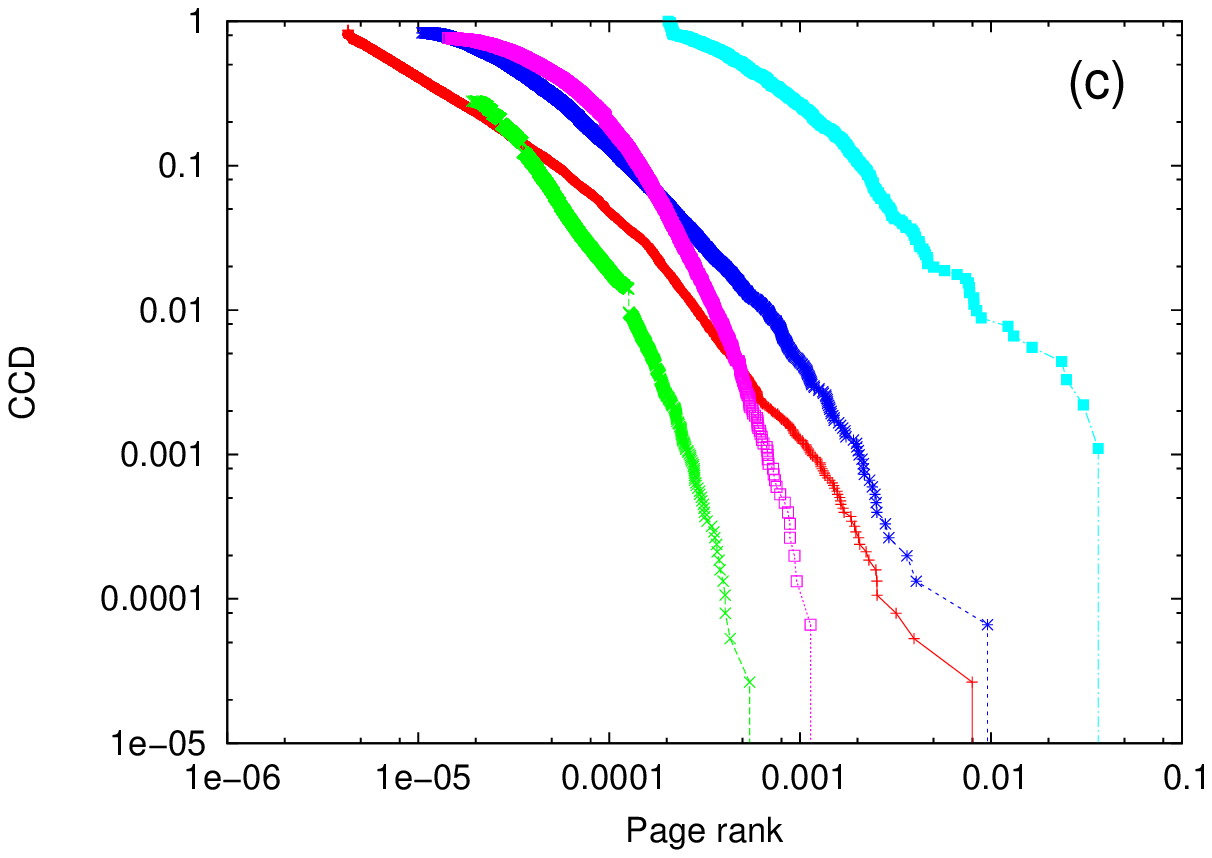}}
\caption{CCD plots for the $y_i$ (a), $x_i$ (b), and $\rho_i$ (c) values of $W$
(Wikipedia), $W'$ (Wikipedia, See also), $M$ (MathWorld), $M'$ (MathWorld, See
also), and $D$ (DLMF).}
\label{fig-searches}
\end{figure}

In Figure~\ref{fig-centralities}, the CCD plots for the $C_i$ and $G_i$ values
share the peculiar property that all nodes are concentrated inside three
relatively narrow centrality intervals. For each of the five libraries, first
are the sink nodes, those for which $R_i=O_i=\emptyset$, having $C_i=G_i=0$.
Then comes what in almost all cases is the most densely populated interval.
Nodes in this interval have the relatively small centrality values typically
associated with relatively large distances to the nodes in $R_i$. They are
members of the graph's largest so-called in-component, which encompasses the
GSCC and all nodes from which at least one directed path leads to the GSCC. This
explains the single exception, which once again concerns the small-GSCC graph of
the Wikipedia library with only See-also links ($W'$). The third centrality
interval contains the remaining nodes and is characterized by centrality values
that in almost all cases bespeak relatively small distances to the nodes in
$R_i$. These nodes lie outside the graph's largest in-component and, once again,
the single exception is relative to $W'$.

\section{Local features and GSCC disruption}\label{local-scc}

In the graphs we have been studying, as in all graphs reflecting real-world
networks, the existence of the GSCC is merely a matter of observation: we simply
look for the graph's strongly connected components and select the largest one.
In a more abstract sense, however, random-graph models of networks have been
studied for the existence of such components under a growth regime from relative
sparseness to relative denseness (that is, as the graph's number of nodes and/or
edges is changed so that it becomes denser). Such studies were initiated with
the Erd\H{o}s-R\'{e}nyi (ER) random graphs \cite{er59}, which are undirected and
characterized by a Poisson distribution of node degrees. Since edges do not have
directions in the ER model, one looks for weakly (rather than strongly)
connected components, or simply connected components, and for the GCC (rather
than the GSCC). It turns out that, increasing $\delta$ (the mean degree) past
$1$ as the graph becomes denser gives sudden rise to the GCC as a connected
component that for the first time is set apart from the others by virtue of its
size \cite{er60}. A similar phenomenon also occurs in many other random-graph
models, including their directed variations with regard to the rise of the GSCC
\cite{k90,mr95,mr98,dms01,nsw01}.

Another similar phenomenon, often called site percolation, is the breakdown of
the GCC or GSCC when nodes are continually isolated from the rest of the graph
by the removal of all edges incident to them. In the case of ER graphs, for
example, the GCC breakdown is expected to happen after a fraction $1-1/\delta$
of the nodes has been randomly isolated \cite{b01}, provided $\delta>1$ to begin
with (i.e., provided there really is a GCC initially). Results of this sort have
been obtained also for undirected graphs with degrees obeying a scale-free
distribution. However, unlike the ER graphs, with their degrees closely
clustered about the mean, now there may exist high-degree nodes, so it makes
sense to look at targeted as well as random node-isolation processes. As it
turns out, for $\alpha=2.5$ (which is thought to describe the Internet graph)
the GCC is only expected to disappear after at least $99\%$ of the nodes have
been randomly isolated, although for relatively small graphs this can be as low
as about $80\%$ \cite{cebh00}. Targeting highest-degree nodes first, though,
implies that isolating fewer than $20\%$ of the nodes is expected to suffice
\cite{ajb00}. We know of no similar studies for directed random-graph models
regarding the impact of node isolation on the graph's GSCC. So, despite the
figures given above, we are essentially left without any meaningful clue as to
what to expect when we conduct node isolation in our five mathematical
libraries.

The results we present in this section describe the evolution of $S$, the
fraction of $n$ inside the GSCC, as nodes are isolated either randomly or
targeting first the non-isolated node for which a specific local feature is
highest. In the former case we provide the average value obtained from ten
independent trials. As for the local feature in question, we report on all ten
discussed in Section~\ref{local}. In all cases, node isolation is performed
until no strongly connected component has more than one node. When isolation
stops, then, all remaining nodes are either isolated (no in- or out-neighbors)
or part of an acyclic portion of the current graph.

Our results appear in Figure~\ref{fig-break}, where the breakdown fractions for
random isolation are seen to be in the $[0.4,0.6]$ interval for the non-See-also
graphs, along with roughly $0.3$ for $M'$ and less than $0.01$ for $W'$. If we
once again except $W'$, with its frail GSCC, and maybe $M'$ as well, we are left
with figures that indicate what seem to be quite resilient GSCCs in $M$, $W$,
and $D$. As we turn to the isolation of nodes following one of the local
features, the data in Figure~\ref{fig-break} reveal that the specific feature in
question is practically irrelevant, with the exception of graph centrality and
closeness centrality in all cases but that of $W'$. These two features are,
respectively, the second and third least effective means we have found to break
the GSCC (following the random method, which is the least effective). Except for
graph and closeness centrality, the data also reveal a breakdown fraction of
about $0.2$ for $W$ and $D$, then a little above $0.1$ for $M$, then a little
below $0.1$ for $M'$ and, finally, less than $0.01$ for $W'$. Targeting nodes
based on any one of these local features, then, reveals a dividing line between
the See-also and non-See-also graphs as well, with $W'$ once more at the lowest
end and $M'$ in between $W'$ and the non-See-also group.

\begin{figure}[p]
\centering
\scalebox{0.65}{\includegraphics{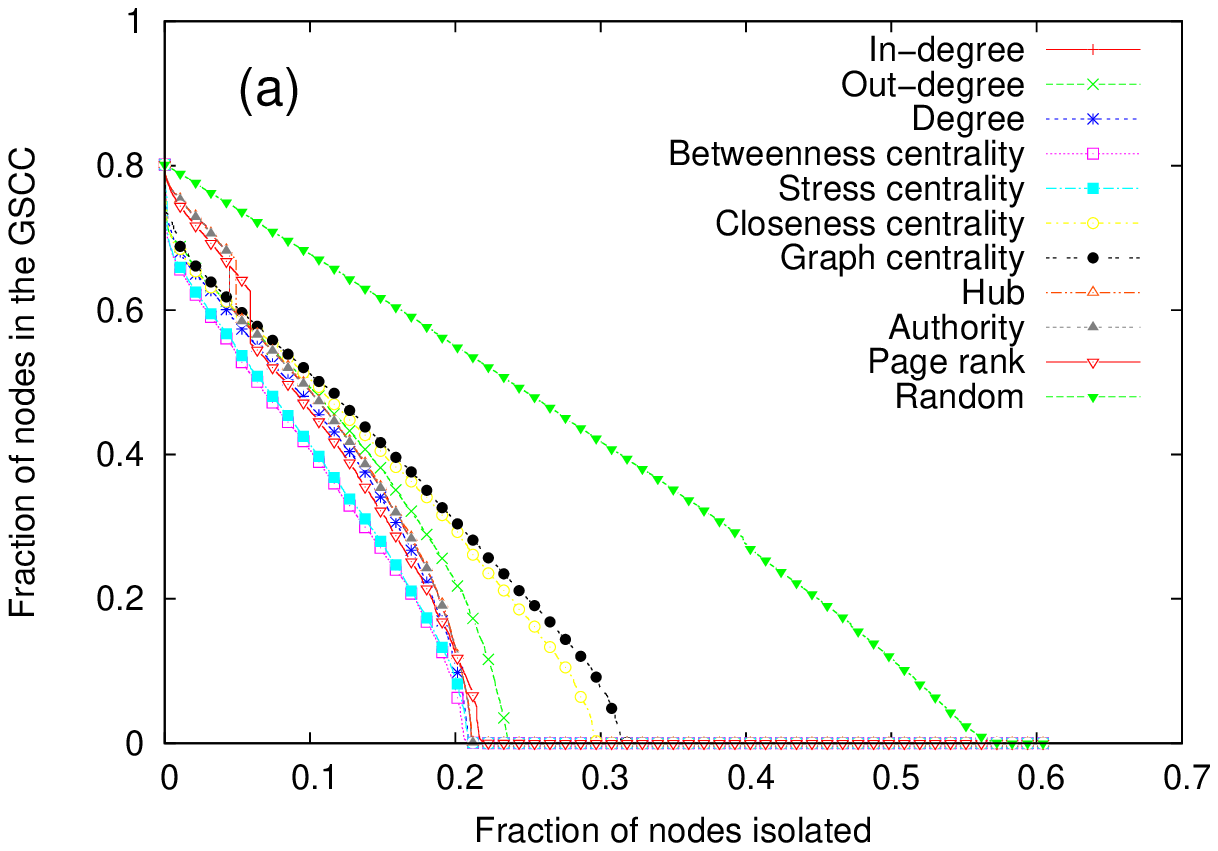}}
\scalebox{0.65}{\includegraphics{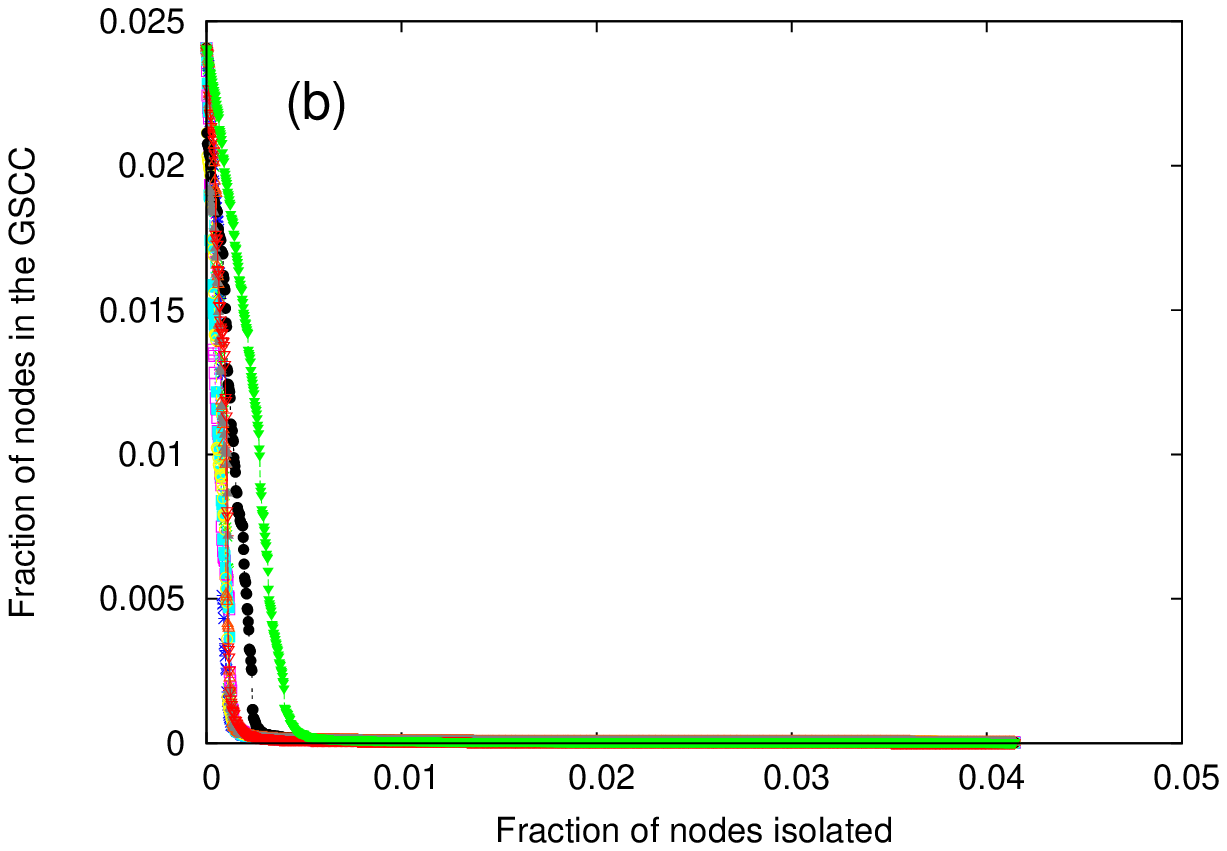}}
\scalebox{0.65}{\includegraphics{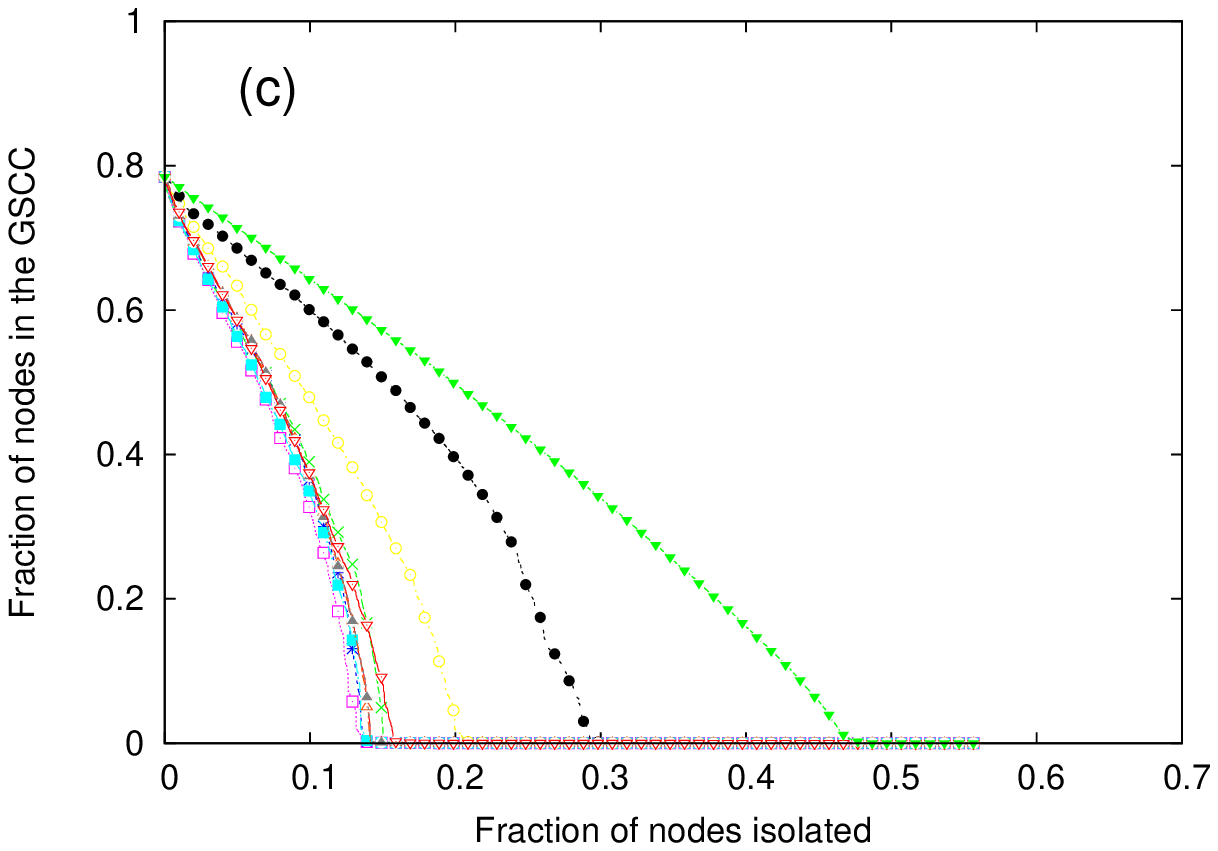}}
\caption{Evolution of $S$ under node isolation in $W$ (Wikipedia; a), $W'$
(Wikipedia, See also; b), $M$ (MathWorld; c), $M'$ (MathWorld, See also; d), and
$D$ (DLMF; e).}
\label{fig-break}
\end{figure}

\addtocounter{figure}{-1}
\begin{figure}[p]
\centering
\scalebox{0.65}{\includegraphics{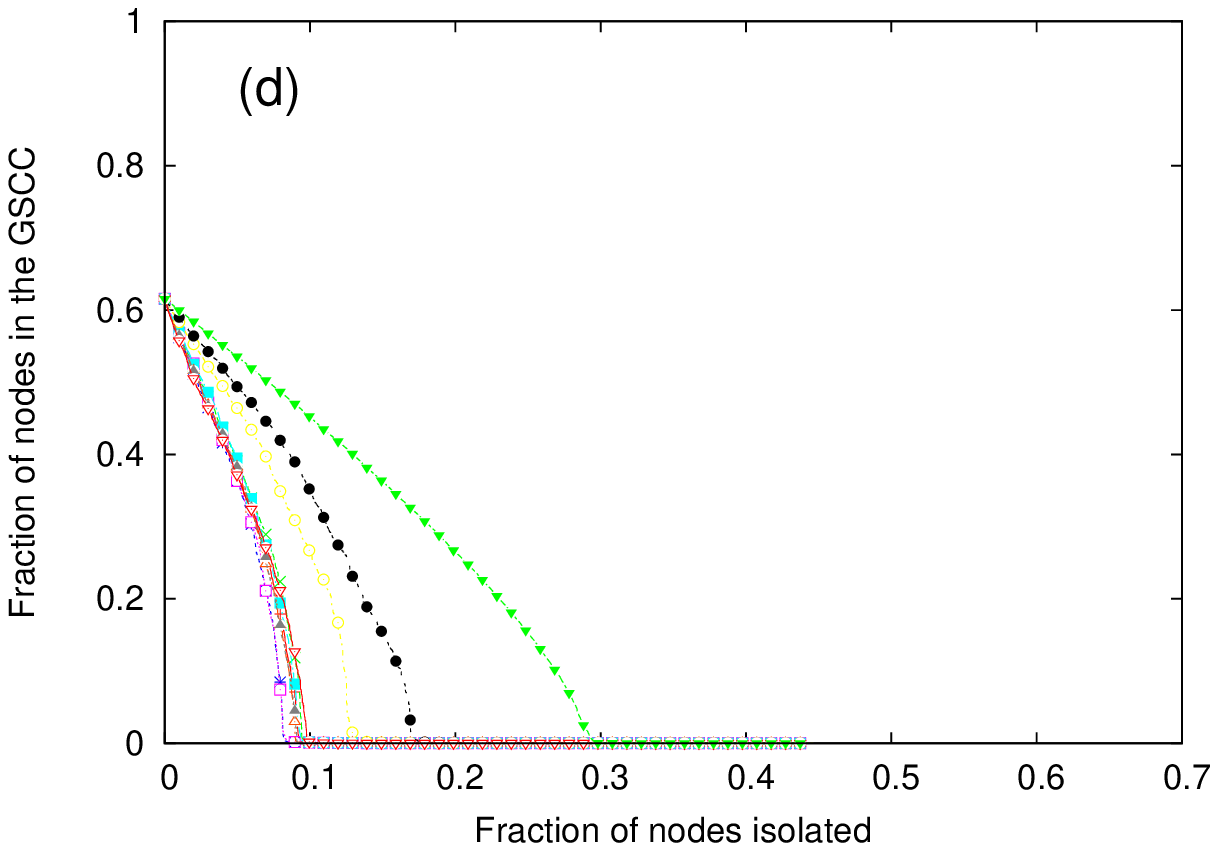}}
\scalebox{0.65}{\includegraphics{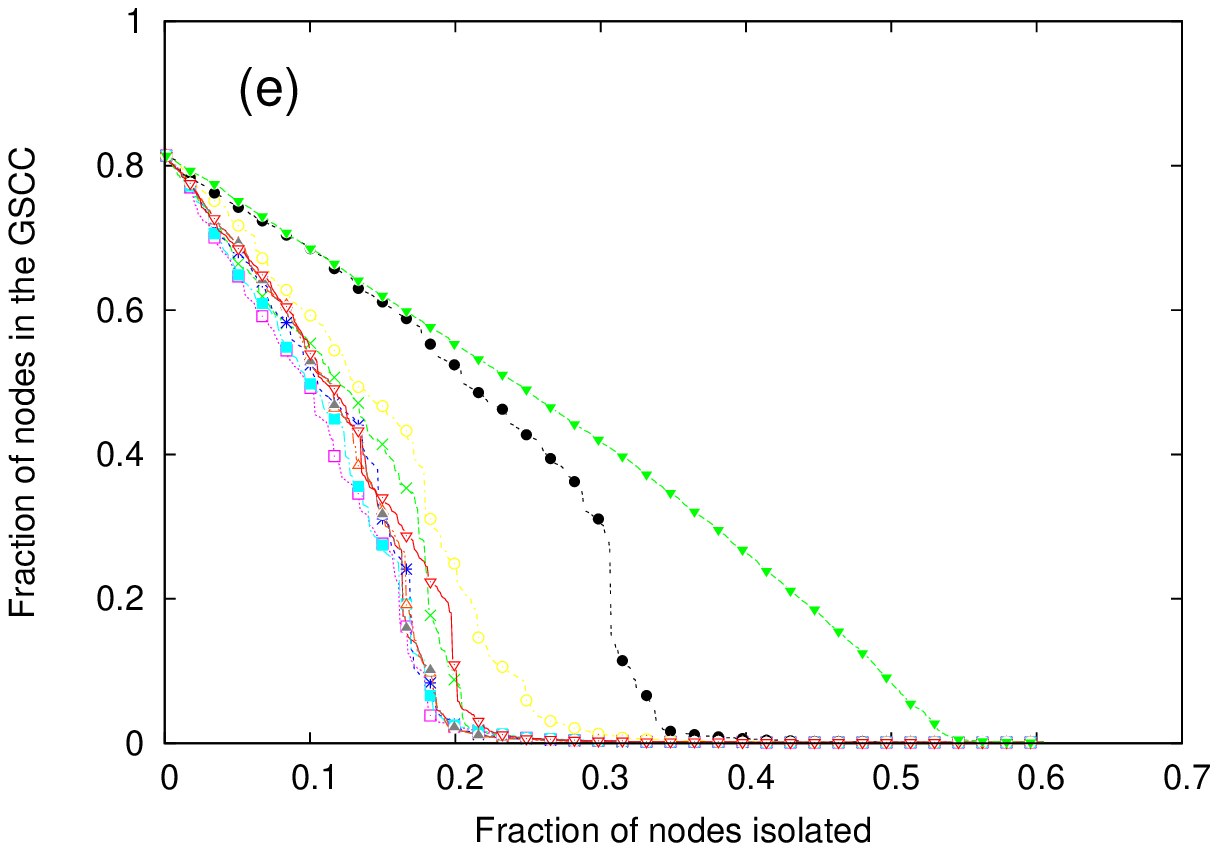}}
\caption{Continued.}
\end{figure}

\section{Local features and text search}\label{local-search}

Google's search engine grew out of the notion of page rank, one of the ten local
features examined in Section~\ref{local}. Page rank, however, is no more of a
node descriptor than any of the other local features, so in principle it is at
least conceivable that any of the others might be used instead. We explore such
possibility in this section for each of the five directed graphs $W$, $W'$, $M$,
$M'$, and $D$, regarding the search, in their nodes' texts, for a number of the
top keywords in mathematics as reported at the Microsoft Academic Search (MAS)
site\footnote{\texttt{http://academic.research.microsoft.com/RankList?entitytype=8\&topDomainID=15\&\linebreak subDomainID=0}.} as of November 1, 2012.

We follow the standard method outlined in \cite{br11}. For each graph and local
feature, and for a given query (one of the aforementioned keywords), this method
begins by identifying a list $A$ of answer nodes (sorted by nonincreasing
feature value) as well as a set $R$ of relevant nodes. It then proceeds to
calculating the well-known Precision and Recall metrics for each
$k=1,2,\ldots,\vert A\vert$. These are given by the fraction of $k$
corresponding to the nodes in the size-$k$ prefix of list $A$ that are also in
$R$ (Precision) and the fraction of $\vert R\vert$ that corresponds to these
shared nodes (Recall). Note that, the more relevant nodes are ranked first in
$A$, the higher Precision values are obtained for a larger stretch of Recall
values.

The elements of $A$ are simply those nodes whose texts contain the keyword in
question. As for $R$, normally it would be identified by a group of experts. In
the absence of one, however, we identify it by resorting to all ten local
features, not just the feature that is being analyzed and was used to sort $A$,
and letting each one ``vote'' for or against each potential candidate for
inclusion in $R$. Set $R$, therefore, is as much a function of the feature
used to sort $A$ as it is of the others. The following steps summarize the
construction of set $R$:
\begin{enumerate}
\item Let $X$ be the set of nodes in whose texts the desired keyword appears. If
$\vert X\vert\le 10$, go to Step~5.
\item Create ten sorted lists of the nodes in $X$, each by nonincreasing order
of one of the ten local features.
\item Let $Y$ be the set of nodes that appear amid the top ten nodes in a strict
majority (i.e., at least six) of the ten lists.
\item Let $R:=Y$ and stop.
\item Let $R:=\emptyset$ and stop.
\end{enumerate}
Note that requiring $\vert X\vert>10$ for termination to occur in Step~4 is
necessary to avoid the trivial case of $R=X$, which allows for no discrimination
of the local features vis-\`{a}-vis one another. When the requirement is not met
and termination occurs in Step~5, the query in question is dropped.

Our results, given next, refer to those MAS keywords, out of the top $300$, for
which the procedure above terminated in Step~4 in our experiments. Whenever such
keywords numbered more than $100$, we considered only the top $100$. As it turns
out, we obtained the desired $100$ keywords for all graphs but $D$, which ended
up with only $14$ keywords (i.e., only $14$ of the $300$ keywords were found in
more than ten nodes). Figure~\ref{fig-pr} contains the resulting
Precision-Recall plots. They are given as Precision averages relative to eleven
Recall intervals, viz.\ $[0,0.1),[0.1,0.2),\ldots,[1,1]$, plotted respectively
at the abscissae $0,0.1,\ldots,1$.

\begin{figure}[p]
\centering
\scalebox{0.65}{\includegraphics{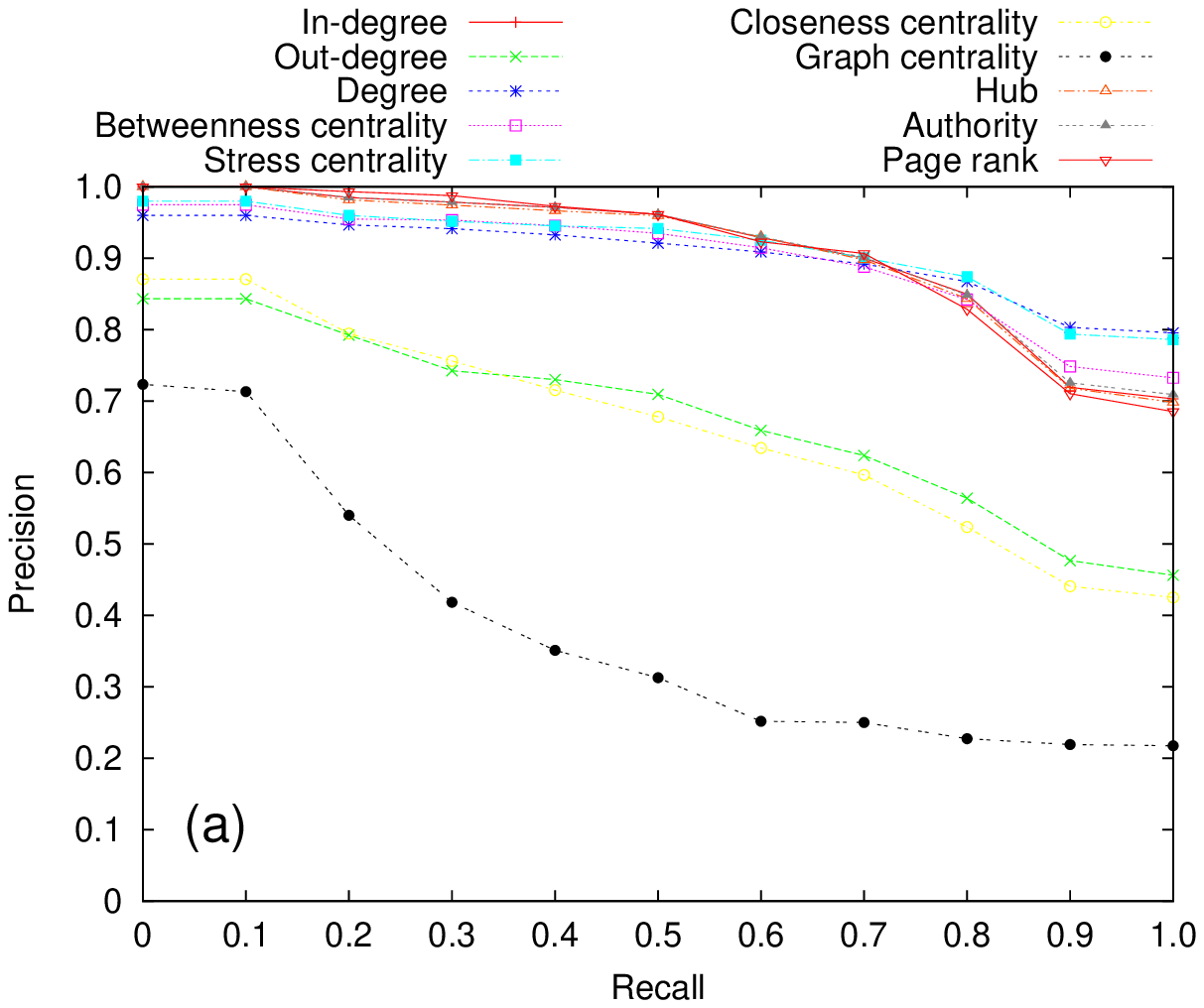}}
\scalebox{0.65}{\includegraphics{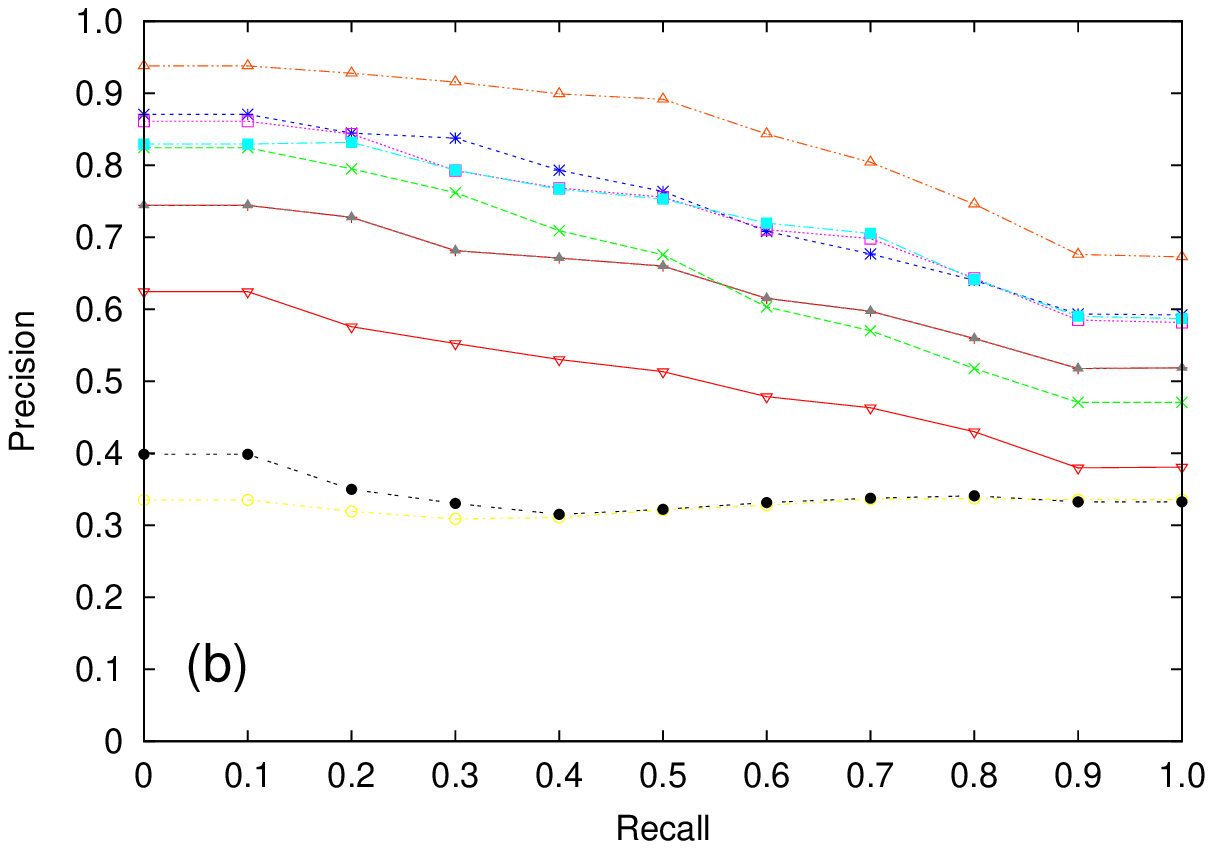}}
\caption{Precision-Recall plots for $W$ (Wikipedia; a), $W'$ (Wikipedia, See
also; b), $M$ (MathWorld; c), $M'$ (MathWorld, See also; d), and $D$ (DLMF; e).}
\label{fig-pr}
\end{figure}

\addtocounter{figure}{-1}
\begin{figure}[p]
\centering
\scalebox{0.65}{\includegraphics{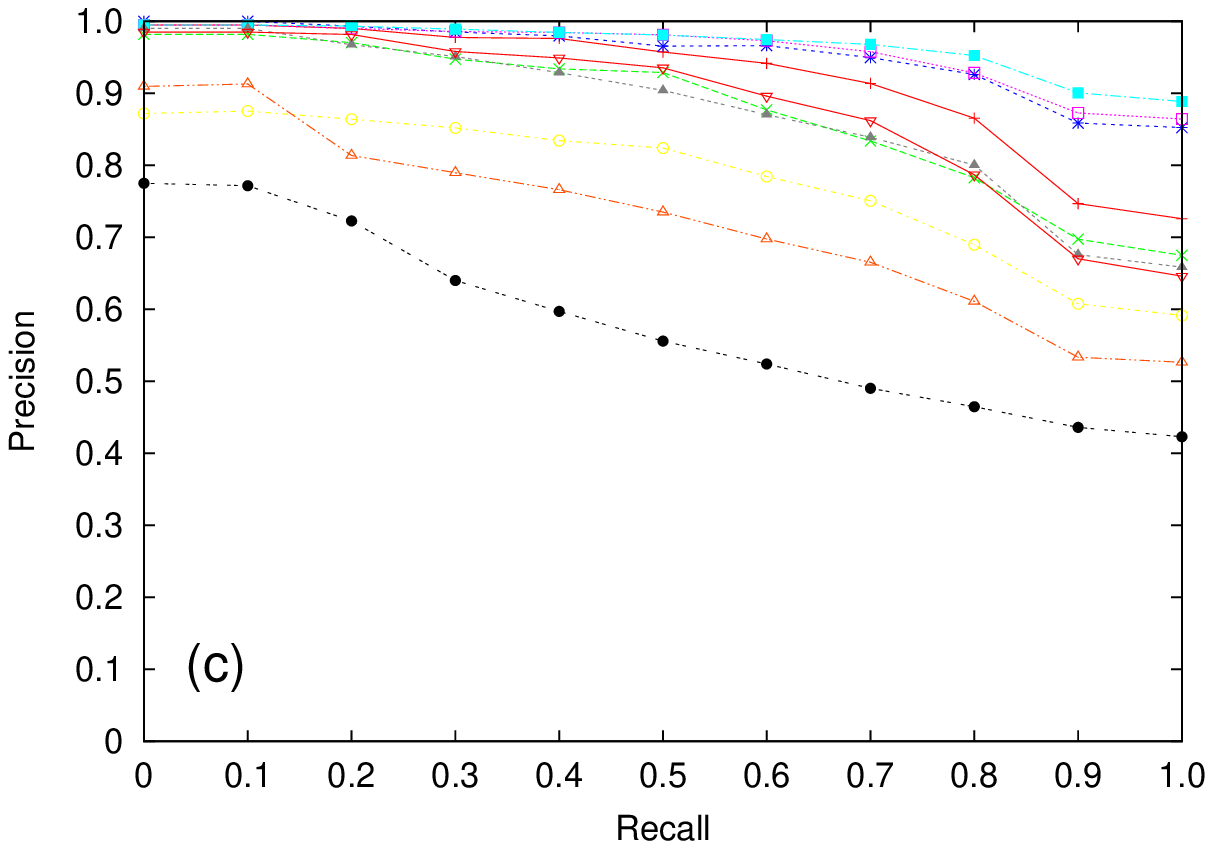}}
\scalebox{0.65}{\includegraphics{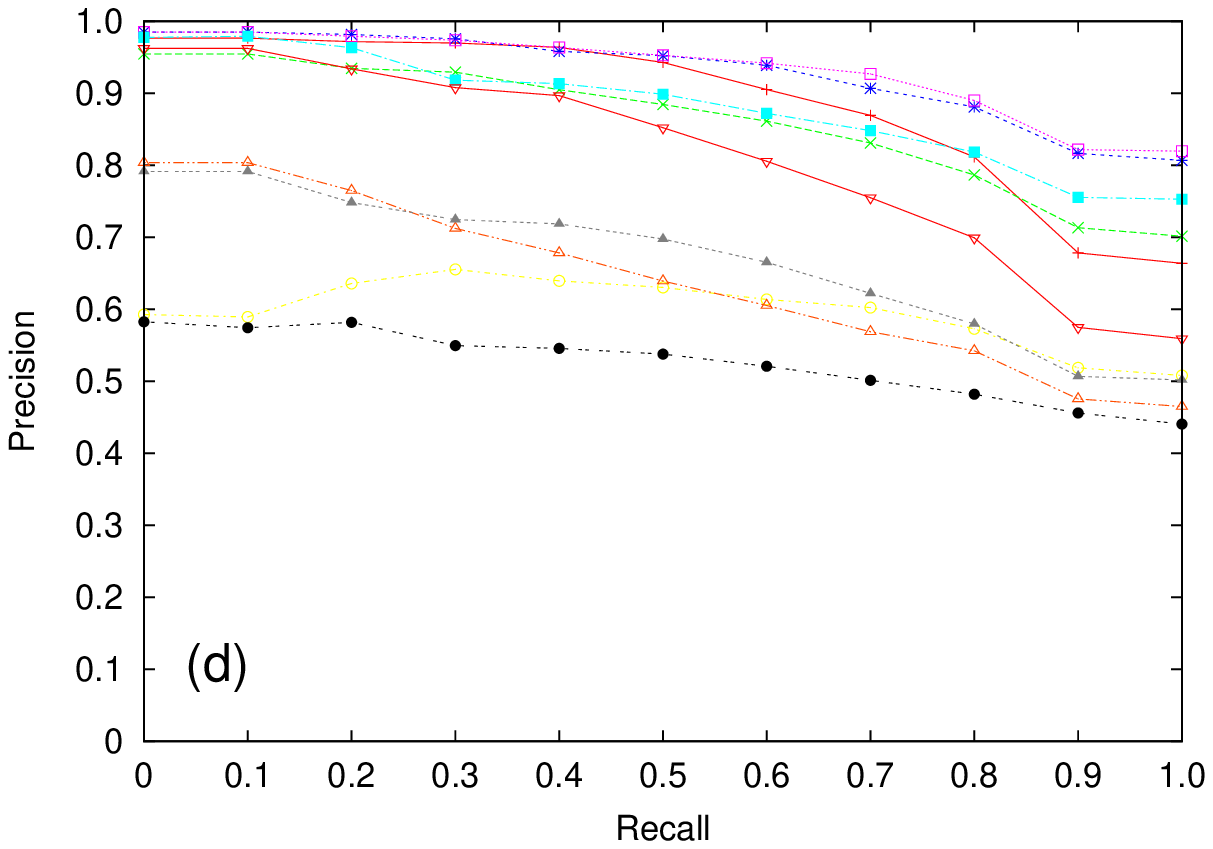}}
\scalebox{0.65}{\includegraphics{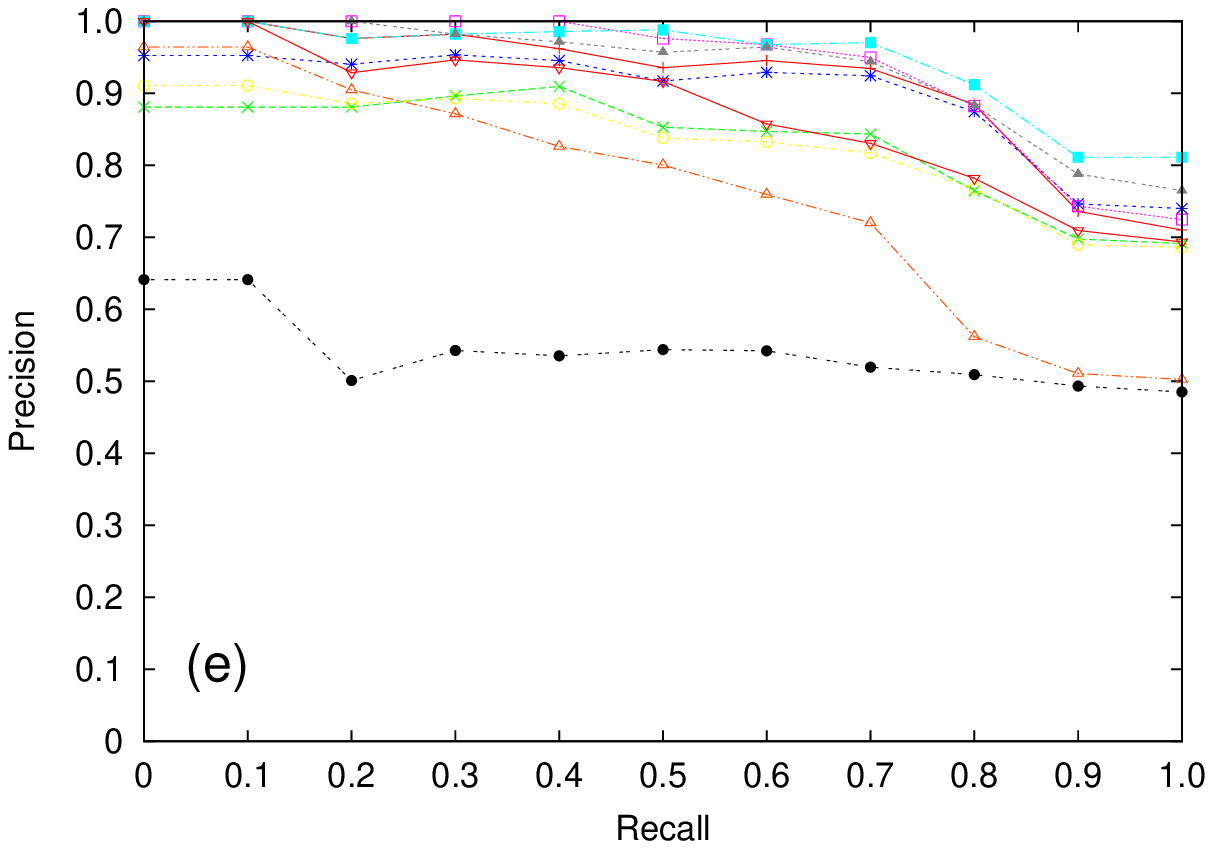}}
\caption{Continued.}
\end{figure}

According to the data in Figure~\ref{fig-pr}, in order to search the
mathematical portion of Wikipedia through the use of local features based on
graph $W$ it is best to use page rank, followed very closely by either the hub
or authority feature. Should the search be based on graph $W'$, however, then
one should use the hub criterion as the absolute champion. Notice,
notwithstanding this, that the use of $W'$ incurs a loss of Precision of about
$10\%$ relative to the use of $W$ and cannot be recommended on any grounds.
Still regarding Wikipedia, our data also indicate that, if one is willing to
examine the list $A$ of answer nodes past the point at which about $70\%$ of the
set $R$ of relevant nodes have been covered, then the nodes' degrees and stress
centralities turn out to be the local features to be preferred for sorting $A$. 

Turning to MathWorld we find a wholly different picture in the data, since now
the best local feature to sort $A$ is the nodes' stress-centrality values as
given by graph $M$, regardless of how much of $R$ one is willing to examine. If
one is willing to examine no more than about $50\%$ of $R$, though, then the
nodes' betweenness-centrality values are equally effective. As for using graph
$M'$, and unlike the case of Wikipedia, only a small loss of Precision is
incurred in comparison with $M$ (about $1$ or $2\%$), but now the local feature
of preference to sort $A$ is betweenness centrality, followed very closely by
the nodes' degrees (for examining up to about $60\%$ of $R$).

As for DLMF, the local feature of choice is once again betweenness centrality
(for up to about $40\%$ of $R$), though the nodes' authority values are equally
effective (up to about $20\%$ of $R$), and so are the nodes' stress-centrality
values and in-degrees (up to about $10\%$ of $R$). Should one be willing to
examine about $50\%$ of set $R$ or more, then stress centrality becomes the
local feature to be preferred.

\section{Conclusions}\label{concl}

We have studied three online mathematical libraries, viz.\ the mathematical
portion of Wikipedia, MathWorld, and DLMF, from the perspective of network
theory. To this end, we considered directed graphs whose nodes are library pages
and whose edges reflect the directed pairing of pages through the links that
point from one to another. In the case of Wikipedia and MathWorld, these links
come in two clearly identifiable categories (those that are in-text and those in
the pages' See-also sections), so we considered two separate graphs for each of
these libraries. We focused on both global and local network-theoretic
properties of these graphs, aiming at characterizing them, studying their
resiliency to the accidental or intentional loss of material, and also assessing
how best to perform text search in the pages that their nodes stand for.

Among our key finds are the presence of GSCCs that in most cases encompass
node fractions substantially larger than that of the Web, indications of
small-world phenomena, practically no signs of relevant assortativity in the
linking patterns, and the absence of any clear power laws describing the
distributions of local features. We also found that most graphs are quite
resilient to the accidental loss of material, though naturally less so when we
consider the intentional destruction of pages. As for searching the libraries
for the occurrence of specific keywords, only for Wikipedia do the customary
criteria of page rank and the HITS-related features perform best. For the
smaller MathWorld and DLMF, primacy is taken by local features that hitherto do
not appear to have been considered for this purpose, notably stress centrality,
betweenness centrality, and the nodes' degrees.

We believe that many of these finds can be attributed to one key distinguishing
property of all three libraries. Unlike what happens in several other domains,
where such intangibles as affinity or popularity dictate the establishment of
connections, in building these libraries what matters is how knowledgeable each
contributor is on the core material being treated and on how it relates to the
other topics. That this key distinction should surface in the form of measurable
effects such as the networks' structural properties and their consequences, and
that this should happen despite the typically large number of often independent
contributors involved, is quite remarkable.

We finalize with a note on some related work on MathWorld that precedes our own
analysis \cite{jg09}. Such work is based on a December 2008 version of the
library, so it predates the one we use by some eight months
(cf.\ Table~\ref{table-graphs}). Despite this relatively short span of
intervening time, our graph has $25\%$ more nodes (about $3\,000$ nodes beyond
that work's $12\,000$), so we conjecture that some intermittent failure during
the download process may have caused the loss of material. In \cite{jg09} the
authors give the distributions of in- and out-degrees and of betweenness
centrality. Despite the considerable difference between the two graphs, our
results agree with theirs in that neither in-degrees nor out-degrees are
distributed as power laws. Their betweenness-centrality distribution also
appears consistent with ours, though they seem to have missed the page for
``Triangle,'' which we find to be one of the top ten for this local feature but
they do not. They also discuss clustering, average distance, and assortativity,
but the definitions they use for these quantities are not the most commonly used
and are incompatible with ours.

\subsection*{Acknowledgments}

The authors acknowledge partial support from CNPq, CAPES, and a FAPERJ BBP
grant.

\bibliography{mlibs}
\bibliographystyle{plain}

\end{document}